\documentclass[preprint,preprintnumbers,amsmath,amssymb,nofootinbib,eqsecnum]{revtex4}
\usepackage{graphicx,xcolor,bm}
\usepackage[linktocpage]{hyperref}

\newcommand{\nn}{\nonumber}
\newcommand{\partialdisplay}[1]{\frac{\partial}{\partial #1}}
\newcommand{\GL}{\operatorname{GL}}
\newcommand{\SO}{\operatorname{SO}}
\newcommand{\D}{\rho}

\newcommand{\NeqFour}{\mathcal N=4}

\def\eqn#1{Eq.~(\ref{#1})}
\def\eqns#1#2{Eqs.~(\ref{#1}) and~(\ref{#2})}
\def\Eqn#1{Eq.~(\ref{#1})}
\def\fig#1{Fig.~\ref{#1}}
\def\figs#1#2{Figs.~\ref{#1} and~\ref{#2}}
\def\sect#1{Section~\ref{#1}}

\def\P{{\rm P}}
\def\NP{{\rm NP}}
\def\tree{{\rm tree}}

\begin{document}

\hfuzz 15pt

\hbox{\hskip0.3cm UCLA/17/TEP/106 \hskip 8.2 cm FR-PHENO-2017-016}
\vskip -.2 cm 

\title{Dual Conformal Symmetry, Integration-by-Parts Reduction,
  \\ Differential Equations and the Nonplanar Sector \\}

\author{Zvi Bern${}^{a}$, Michael Enciso${}^{a}$, Harald Ita${}^{b}$ and Mao Zeng${}^{a}$}
\affiliation{
\vskip .2 cm 
${}^a$Mani L. Bhaumik Institute for Theoretical Physics\\
Department of Physics and Astronomy\\
University of California at Los Angeles\\
Los Angeles, CA 90095, USA\\
\vskip -5 mm
${}^b$Physikalisches Institut\\
Albert-Ludwigs-Universit\"at Freiburg\\
D–79104 Freiburg, Germany\\
}


\begin{abstract}
We show that dual conformal symmetry, mainly studied in planar
$\mathcal N=4$ super-Yang--Mills theory, has interesting consequences
for Feynman integrals in nonsupersymmetric theories such as QCD,
including the nonplanar sector. A simple observation is that dual
conformal transformations preserve unitarity cut conditions for any
planar integrals, including those without dual conformal
symmetry. Such transformations generate differential equations without
raised propagator powers, often with the right hand side of the system
proportional to the dimensional regularization parameter $\epsilon$.
A nontrivial subgroup of dual conformal transformations, which leaves
all external momenta invariant, generates integration-by-parts
relations without raised propagator powers, reproducing, in a simpler
form, previous results from computational algebraic geometry for
several examples with up to two loops and five legs. By opening up the
two-loop three- and four-point nonplanar diagrams into planar ones, we
find a nonplanar analog of dual conformal symmetry.  As for the planar
case this is used to generate integration-by-parts relations and differential equations.  This implies
that the symmetry is tied to the analytic properties of the nonplanar
sector of the two-loop four-point amplitude of $\mathcal N = 4$
super-Yang--Mills theory.
\end{abstract}

\maketitle
\tableofcontents

\section{Introduction}
\label{sec:intro}

Dual conformal symmetry is a hidden symmetry of planar $\mathcal N =4$
super-Yang-Mills theory~\cite{ConformalIntegrals,DCI} which puts
strong constraints on the analytic structure of its scattering
amplitudes.  In this paper we will discuss applications of this
symmetry towards questions of practical interest in generic theories,
such as finding useful and compact integration-by-parts (IBP)
relations and differential equations (DEs) for loop integrals.  
We also use these ideas to extend the symmetry to the
nonplanar sector by explicitly constructing it for the full two-loop
four-point amplitude of $\mathcal N = 4$ super-Yang--Mills theory.  As
for the planar case, the symmetry leads to useful IBP
relations and DEs.

An important feature of the IBP relations and DEs generated by dual
conformal transformations is that they are naturally compatible with
generalized unitarity~\cite{GeneralizedUnitarity}, which is a powerful
method for computing multi-loop scattering amplitudes.  Generalized
unitarity helps to overcome the fast growth of complexity as the loop
order and the number of legs increase.  At one loop,
unitarity-compatible integrand-based
reduction~\cite{OPP,Forde,BlackHat} simplifies loop amplitudes to a
linear combination of master integrals, with coefficients determined
from generalized unitarity cuts.  This has led to tremendous progress,
including the ``NLO revolution'' for computing NLO QCD corrections for
collider processes (see e.g. Refs.~\cite{NLOExamples}).  To extend the
reach of generalized unitarity to generic theories at higher loops, it
is natural to retain the following two important properties: (i) the
parameterization is minimal without redundant parameters, leading to
invertible linear systems which can be solved to determine the
integrand; (ii) the integrand is decomposed into master integrands and
spurious integrands that vanish upon integration, so only the
coefficients of the master integrands are needed to evaluate the
amplitudes.

These methods for evaluating scattering amplitudes offer great promise
to tackle general problems at two loops and beyond (see
e.g. Ref.~\cite{BH4DGluon}).  For dimensionally regularized integrals
beyond one loop, it is in fact easy to write down a parameterization
that satisfies property (i) by identifying a minimal set of
``irreducible numerators'' that cannot be expressed as linear
combinations of inverse propagators.  For integrals in integer (most
often four) dimensions, the problem is more intricate, as Gram
determinant identities further reduce the number of independent terms
in the integrand. But a complete and computationally efficient
solution has been found using polynomial division
algorithms~\cite{NumPar,NumParM}.  To construct a parameterization to
satisfy the above property (ii), a first step has been developed in
the mentioned papers exploiting the rotation symmetry in the
``transverse'' directions orthogonal to all external momenta.  This is
in direct analogy with the one-loop case~\cite{OneLoopEKS}. A second
step, which is substantially more nontrivial, is to identify all
remaining contributions that integrate to zero.  At higher loops the
only known practical means to accomplish this~\cite{ItaIBP} is to
exploit IBP relations~\cite{IBP} without increasing propagator
powers~\cite{KosowerNoDouble}, to not only simplify the problem, but
to make it naturally compatible with generalized unitarity.  Our
approach based on exploiting dual conformal transformations
automatically generates IBP relations with these properties.

Geometrically, the special IBP relations which do not lead to higher
propagator powers are generated by polynomial vector fields that are
tangent to the unitarity cut surface~\cite{ItaIBP}.  This is related
to the tangent algebra studied in the mathematics
literature~\cite{HauserMullerMath}, as pointed out in
Ref.~\cite{ZhangAlgGeom}. A key problem for generating
unitarity-compatible IBP relations is finding these special
IBP-generating vectors. One solution is to solve ``syzygy equations''
using computational algebraic geometry~\cite{KosowerNoDouble,
  Schabinger, LarsenZhang, BH4DGluon}. This is often time-consuming
for the more complicated multi-loop integrals, and produces lengthy
and unenlightening results. Analytic insights into the IBP-generating
vectors from Ref.~\cite{ItaIBP} shows that for generic two-loop
integrals with massive external legs and internal propagators, a
complete set of IBP-generating vectors comes from simple combinations
of one-loop rotation vectors. These vectors will be referred to as
``generic" vectors, and can be constructed as minors of matrices in
Section \ref{sec:Landau}. For Feynman integrals involving vanishing or
degenerate mass configurations, however, ``exceptional" IBP-generating
vectors appear, in addition to the generic vectors found by the
aforementioned reference. This leads to extra IBP relations,
e.g.\ relations between one-loop triangle integrals and bubble
integrals.  A full analytic understanding of these exceptional
vectors is still missing in the literature, though a connection with
singularities of unitarity cut surfaces has been explored
\cite{ZhangAlgGeom}. We will show that important missing insights, at
least for the integral topologies covered in this paper, come from
$\mathcal N=4$ super-Yang--Mills theory.

In the study of scattering amplitudes, theories with more symmetries
have often led to unexpected simplifications for theories with fewer
symmetries. For example tree-level gluon amplitudes in pure Yang-Mills
have hidden supersymmetry because they coincide with the same
amplitudes in super-Yang--Mills theory~\cite{ParkeTaylorSusy}.  A
one-loop example is that supersymmetric decompositions can be applied
to nonsupersymmetric theories~\cite{LoopSusy,GeneralizedUnitarity}.
Following this philosophy, we aim to develop a relatively simple
analytic understanding of IBP-generating vectors for a variety of one-
and two-loop Feynman integrals with vanishing or degenerate masses,
using dual conformal symmetry of planar $\NeqFour$ super-Yang--Mills
theory as a guiding principle.  The use of dual conformal symmetry
also extends to a large class of planar Feynman integrals in even
integer dimensions, with an appropriate number of
propagators~\cite{ConformalIntegrals, CrossRatios,
  CaronHuotEmbedding}. This is easiest to implement for planar
diagrams where dual conformal symmetry is defined, but as we shall see
by opening up nonplanar diagrams into planar diagrams~\cite{NonplanarYangian}, we identify a symmetry that is
analogous to dual conformal symmetry.

When we consider integrals in arbitrary dimensions, generic numerators
or integrals with too few propagators, the symmetries are lost because
the numerators cannot balance the conformal weights from the
denominators and the integration measure.  However, for our purpose of
finding IBP-generating vectors, only the geometry of the unitarity cut
surface, fixed by the propagators not the numerators, is
relevant. Therefore we can still find insights from dual conformal
symmetry in order to analyze the loop integrals of any theory more
generally. It turns out that a subgroup of dual conformal
transformations, which leaves external momenta unchanged, generates
infinitesimal shifts in the loop momenta to produce IBP relations
without higher-power propagators.  This is connected to the fact that
under dual conformal transformations and their nonplanar
generalization, the infinitesimal variations of inverse propagators
are proportional to the inverse propagators themselves. It turns
out that IBP-generating vectors obtained from conformal
transformations contain exceptional vectors which we seek to
understand. In the process we also find that at one loop the
exceptional vectors relate directly to Landau
equations~\cite{Landau,SMaxrixBooks}.

To illustrate the ideas in a simple context, we first present a number
of one-loop examples. As a toy example we illustrate the case of the
one-loop triangle diagram with a single external mass. While standard
integral reductions~\cite{PV,OPP,Forde} reduce tensor triangle
integrals to the scalar triangle integral, we show that dual conformal
transformations can be directly applied to reduce the scalar triangle
integral to bubble integrals. Then we use this example to illustrate 
the embedding formalism~\cite{SimmonsDuffin,CaronHuotEmbedding} which
reduces conformal transformations in an $\SO(d-1,1)$ dual spacetime to
linear Lorentz transformations in an $\SO(d,2)$ embedding space.  The
latter treatment will involve a general algorithm that can be applied
to all one-loop integrals. Finally, we turn to two-loop examples,
including nonplanar cases. We adopt a level-by-level approach to IBP
reduction. For each topology, we only identify IBP relations which
reduce all tensor integrals to top-level master integrals and
lower-level integrals with fewer propagators. One can descend into the
lower-level topologies recursively to accomplish the complete IBP
reduction.

We also use dual conformal symmetries to generate DEs
for integrals~\cite{DEs}.  This has proven to be a powerful means for
evaluating integrals. The DEs we generate are in
terms of integrals without propagators raised to higher powers, along
the lines of Ref.~\cite{DEnoDouble}. For the integrands that would be
invariant in four dimensions under dual conformal transformations or
their nonplanar analogs, the right hand side of the DEs
are  automatically proportional to the dimensional
regularization parameter $\epsilon = (4-d)/2$.  If there were no
infrared singularities, we could take $\epsilon \rightarrow 0$, and
the right side of the DEs would vanish.  This
property is already known for such integrals, after reducing
to a carefully chosen basis of integrals~\cite{HennDifferentialEqs, Henn2013nsa}.
In our case, it follows from the existence of a symmetry.

Besides the practical utility of IBP relations and DEs, our
considerations point to a nontrivial generalization of dual conformal
symmetry to the full nonplanar sector of $\mathcal N =4$
super-Yang--Mills theory.
Refs.~\cite{TwoLoopN4Nonplanar,NonplanarAmplituhedron} found in a
variety of nontrivial examples that the analytic properties implied by
dual conformal symmetry such as having only logarithmic singularities,
no poles at infinity and other properties carry over to the nonplanar
sector.  What symmetries might be behind this?  In this paper we take
initial steps towards understanding the symmetries behind these
properties, by building on the connection between dual conformal
transformations and polynomial tangent vectors of unitarity cut
surfaces.  For the case of the nonplanar sector of the two-loop
four-point amplitude~\cite{BDDPR} based on our analysis of symmetries
of integrals we show that there is indeed a symmetry analogous to dual
conformal symmetry.

This paper is organized as follows.  In Section \ref{sec:key}, we
review unitarity-compatible IBP relations, dual conformal
transformations and the embedding formalism which linearizes the
transformations.  In Section \ref{sec:triangle}, we illustrate the
application of dual conformal transformations, starting from the
simple toy example of the one-loop triangle with massless propagators
and one massive external leg. Two parallel treatments are presented,
one based directly on dual conformal transformations in $d$ dimensions
and the other on the $\SO(d,2)$ embedding space. The latter part of
the section will present two more complicated examples at one
loop, namely the triangle diagram with two external masses, and the
massive triangle diagram involved in QCD corrections of the
$H\rightarrow b \bar b$ decay.  Section \ref{sec:planarTwoLoop} gives
two-loop planar examples, reproducing nontrivial IBP-generating
vectors previously obtained from computational algebraic geometry.
Section~\ref{sec:DE} outlines applications to obtain DEs using
transformations that act nontrivially on the external momenta.
Section~\ref{sec:nonplanar} formulates a nonplanar analog of dual
conformal symmetry.  Applications to IBP and differential equations
for dimensionally-regularized nonplanar integrals are also worked out.
In Section~\ref{sec:nonplanaramplitude}, we show the invariance of the
two-loop four-point amplitude of $\mathcal N=4$ super-Yang--Mills
theory under this symmetry.  Finally in Section~\ref{sec:Landau} we
describe the interesting connection between the exceptional IBP 
vectors and Landau equations.  Our conclusions and outlook are
presented in Section~\ref{sec:Conclusion}.  An appendix giving
matrices describing the dual conformal transformations of the two-loop
pentabox integrals is also included.

\section{Basic concepts}
\label{sec:key}

In this section we give an overview of basic concepts that will be
useful for the remainder of the paper.  We first review the notion of
unitarity-compatible IBP relations that do not increase the propagator
powers, which generically occurs whenever derivatives hit propagators.
Then we discuss using dual conformal transformations as a means for
generating IBP relations that are compatible with unitarity cuts and do
not increase the powers of the propagators.  We will also review the
embedding formalism for dual conformal transformations. This will be
useful in subsequent sections, since it reduces conformal
transformations to simpler Lorentz transformations in two higher
dimensions.

\subsection{Unitarity-compatible IBP relations}
\label{subsec:UCIBP}
Consider an $L$-loop Feynman integral with $L$ independent loop
momenta, $l_1, l_2, \dots, l_L$, $M$ external legs with momenta $p_i$,
$1 \leq i \leq M$, and $N$ propagators, $1/\D_j$, $1 \leq j \leq N$,
\begin{equation}
\int \prod_{A=1}^L d^d l_A \frac{\mathcal N}{\prod_j \D_j} \,.
\label{eq:feyn}
\end{equation}
where $\mathcal N$ is a numerator that has polynomial dependence on
all possible Lorentz-invariant dot products amongst loop and external
momenta.

Integration-by-parts relations~\cite{IBP} arise because total
derivatives integrate to zero in dimensional regularization,
\begin{equation}
0 = \int \prod_{A=1}^L d^d l_A \, \partialdisplay{l_B^\mu} 
  \frac{v_B^\mu \, \mathcal N}{\prod_j \D_j}\,, 
\label{eq:ibp}
\end{equation}
where there is implicit summation over the loop momentum label $B$,
and $v_B^\mu$ is built out of all possible Lorentz vectors $p_i^\mu$
and $l_A^\mu$, each multiplied by polynomials in Lorentz-invariant dot
products. The identity amongst integrals comes from explicitly applying
the derivative. We will refer to
\begin{equation}
    v_B^\mu \partialdisplay{l_B^\mu}\,,  \label{eq:vector}
\end{equation}
as an IBP-generating vector or IBP vector.

If the vector satisfies the condition~\cite{KosowerNoDouble},
\begin{equation}
v_B^\mu \partialdisplay{l_B^\mu} \D_j = \mathcal W_j \D_j \,, 
\label{eq:tangent}
\end{equation}
where there is an implicit sum over $B$ and $\mu$, for each $1 \leq j
\leq N$, with the $\mathcal W_j$ being polynomials in Lorentz-invariant
dot products, then the IBP relation \eqn{eq:ibp} will not lead to
propagators raised to two or more powers. More generally speaking, if
we start with some propagator raised to a power, the power of that 
propagator will not be increased further in the IBP relation
\cite{BH4DGluon}. This will be called a ``unitarity-compatible'' IBP
relation, as unitarity cut conditions are easily imposed when there
are no raised propagator powers.  The standard ways to find IBP
vectors that satisfy \eqn{eq:tangent} are based on solving syzygy
equations~\cite{KosowerNoDouble,Schabinger,LarsenZhang,BH4DGluon},
often using software for computational algebraic
geometry~\cite{SyzygySoftware}.

This is natural with the unitarity approach.  If a certain inverse
propagator $\D_j$ is set to zero by a unitarity cut, then for that
case the right hand side of \eqn{eq:tangent} is zero, which means the
IBP-generating vector is a tangent vector to the unitarity cut surface
of \emph{any} cut, maximal or non-maximal ~\cite{ItaIBP}. It should be
emphasized that it is a polynomial (rather than rational) tangent
vector.

\subsection{Unitarity-compatible differential equations}
\label{subsec:UCDE}

A powerful method for evaluating Feynman integrals is differential
equations with respect to external momenta~\cite{DEs}. In this method,
one computes derivatives
\begin{equation}
\chi_i^\mu \partialdisplay{p_i^\mu} \int \prod_{A=1}^L d^d l_A \frac{\mathcal N}{\prod_j \D_j} \,, 
\label{eq:diffFeyn}
\end{equation}
where there is implicit summation of $i$ over every external momentum,
and $\chi_i^\mu$ generates an infinitesimal change in the kinematic
invariants (i.e.\ Lorentz-invariant dot products between external
momenta). We require $\chi_i^\mu$ to have no dependence on loop momenta. 
Since total derivatives vanish upon integration,
Eq.~\eqref{eq:diffFeyn} is equivalent to
\begin{align}
& \int \prod_{A=1}^L d^d l_A \left[
\chi_i^\mu \partialdisplay{p_i^\mu} \frac{\mathcal N}{\prod_j \D_j} + \partialdisplay{l_B^\mu} \frac{v_B^\mu \, \mathcal N}{\prod_j \D_j} \right] \nn \\
& \hskip 3 cm \null =  \int \prod_{A=1}^L d^d l_A \left[ \frac{\partial v_B^\mu}{\partial l_B^\mu} + \left( \chi_i^\mu \partialdisplay{p_i^\mu} + v_B^\mu \partialdisplay{l_B^\mu} \right) \right] \frac{\mathcal N}{\prod_j \D_j} \, .
\label{eq:DEformula}
\end{align}
We will refer to
\begin{equation}
\chi_i^\mu \partialdisplay{p_i^\mu} + v_B^\mu \partialdisplay{l_B^\mu}
 \label{eq:DEvec}
\end{equation}
as the DE-generating vector. Under the condition~\cite{DEnoDouble}
\begin{equation}
\left( \chi_i^\mu \partialdisplay{p_i^\mu} + v_B^\mu \partialdisplay{l_B^\mu} \right) \D_j = \mathcal W_j \D_j \, , \label{eq:tangentDE}
\end{equation}
for some polynomial $\mathcal W_j$ for each $1 \leq j \leq N$,
Eq.~\eqref{eq:DEformula} has no propagators raised to higher powers,
i.e.\ is unitarity-compatible. In our framework, IBP-generating
vectors are special cases of DE-generating vectors without external
momentum derivatives. Similarly, IBP relations are regarded as special
cases of differential equations whose left hand side is zero rather
than an external momentum derivative of the integral. Similar to the interpretation of Eq.~\eqref{eq:tangent}, Eq.~\eqref{eq:tangentDE} implies that the DE-generating vector is a tangent vector to unitarity cut surfaces, considered as solutions to unitarity cut conditions in the space of \emph{both} external and loop momenta.

We will refer to $\mathcal W_j$ as the weight of the inverse
propagator $\rho_j$ under the infinitesimal transformation of
$p_i$ and $l_B$ generated by the vector
\eqref{eq:DEvec}. The total divergence term $\partial v_B^\mu /
\partial l_B^\mu$ in Eq.~\eqref{eq:DEformula} may be regarded as the
weight $\mathcal W_{\rm measure}$ of the integration measure, coming
from an infinitesimal deviation of the Jacobian from unity (see a
later discussion around Eq.~\eqref{eq:weightOfDz}), under the
same transformation. In addition, in some cases of interest, the
numerator $\mathcal N$ also has a well-defined weight $\mathcal
W_{\mathcal N}$ with polynomial dependence on external and loop
momenta. In this case Eq.~\eqref{eq:DEformula} is rewritten as
\begin{align}
&\quad \int \prod_{A=1}^L d^d l_A \left[ \frac{\partial v_B^\mu}{\partial l_B^\mu} 
+ \left( \chi_i^\mu \partialdisplay{p_i^\mu} + v_B^\mu \partialdisplay{l_B^\mu} \right) \right] \frac{\mathcal N}{\prod_j \D_j} \nn \\
& \hskip 3 cm \null = \int \prod_{A=1}^L d^d l_A \left( \mathcal W_{\rm measure} + \mathcal W_{\mathcal N} 
- \sum_k \mathcal W_k \right) \frac{\mathcal N}{\prod_j \D_j} \,. 
\end{align}
If in the above equation,
\begin{equation}
\mathcal W_{\rm measure} + \mathcal W_{\mathcal N} - \sum_k \mathcal W_k = 0 \,,
\end{equation}
then the integral is formally invariant under the infinitesimal
transformation generated by the vector~\eqref{eq:DEvec}. A
trivial example is a Lorentz transformation (in both external and loop
momenta), under which the integration measure, propagators, and the
numerator are separately invariant.  In most cases the integrals
are infrared singular and an infrared regulator is needed.  This
shifts the weight of the measure factor by terms proportional to
$\epsilon$, making the symmetry anomalous.  

\subsection{Properties of IBP- and DE-generating vectors}
IBP-generating vectors defined by Eq.~\eqref{eq:tangent} and DE-generating
vectors defined by Eq.~\eqref{eq:tangentDE} satisfy the following
properties:

First, if an IBP-generating vector (or DE-generating vector) is
multiplied by a polynomial in Lorentz-invariant dot products of
external and loop momenta, it is still a valid IBP-generating
vector (or DE-generating vector). Furthermore, the linear combination
of two IBP-generating vectors (or DE-generating vectors) is still a
valid vector. Therefore, IBP- and DE-generating vectors form
\emph{modules} over the ring of polynomials.

Second, by applying Eq.~\eqref{eq:tangentDE} twice, it can be seen
that the composition of two DE-generating vectors still does not raise
the power of any propagator. Furthermore, the components $\chi_i^\mu$
remain independent of the loop momenta.  This can be used to compute
higher-order differential equations~\cite{Laporta2004rb, MullerStach2011ru} without generating
doubled propagators.

Third, it follows from the second property above that IBP- and
DE-generating vectors form a closed Lie algebra. The action of the DE
vector Eq.~\eqref{eq:DEvec} in Eq.~\eqref{eq:DEformula} is, in the
language of differential geometry, the Lie derivative action on the
form
\begin{equation}
\prod_{A=1}^L d^d l_A \, \frac{\mathcal N}{\prod_j \D_j} \, .
\end{equation}
It is well known that the Lie derivative action of vectors commutes
with the Lie bracket of vectors, i.e.\ the Lie algebra structure
extends to the action of IBP- and DE-generating vectors. This is
essentially the observation of Ref.~\cite{RomanLeeGroupStructure} in
the slightly different context of IBP reduction with doubled
propagators. As in the aforementioned reference, the Lie algebra
structure allows us to reduce the redundancy of IBP relations---all
the necessary IBP relations arise from the action of a minimal
generating set of IBP vectors on the possible tensor integrals.

Fourth, given the unitarity-compatible conditions
Eq.~\eqref{eq:tangent} and \eqref{eq:tangentDE}, the IBP- and
DE-generating vectors are valid on unitarity cuts and can be used to
generate relations between cut integrals~\cite{CutIntegrals}.

\subsection{Dual conformal symmetry}

If the Feynman integral \eqn{eq:feyn} is planar and only has
massless propagators, we can write each inverse propagator as either
\begin{equation}
(y_A - y_B)^2, \hskip 1 cm (A \neq B) \,, \label{eq:la-lb}
\end{equation}
or
\begin{equation}
(y_A - x_j)^2 \,, \label{eq:la-qj}
\end{equation}
where $A$ and $B$ are loop-momentum labels, and $x_j$ are the vertices
of a coordinate-space polygon whose edge $(x_{i+1}-x_i)$ is equal to the
external momenta $p_i$. We will refer to $x_j$ as external momentum
points and $y_A$ as loop-momentum points. 
This is known as the dual-space version of planar Feynman integrals,
as each $y_A$ and $x_j$ may be considered as coordinate-space points
in a dual $\SO(d-1, 1)$ ``spacetime'' (not to be confused with ordinary spacetime). 

\begin{figure}
  \centering
  \includegraphics[width=0.5\textwidth]{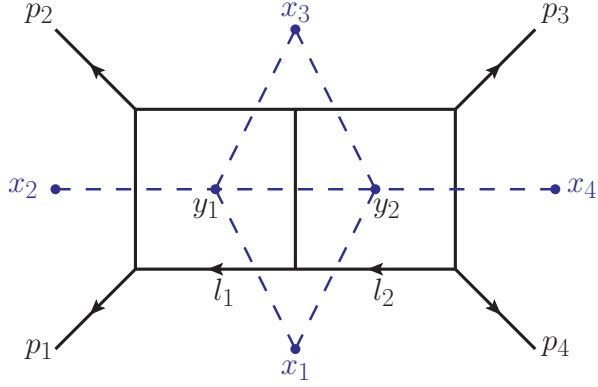}
  \caption{The double box integrals.  Differences of the dual points
    give momenta flowing in the diagram.  The $x_i$ and $y_i$ are dual
    coordinates the double box.  The dual diagram is given by the
    dashed (blue) diagram.  }
  \label{fig:dbox-dual}
\end{figure}

As a simple example, consider the two-loop planar double-box integral, 
\begin{equation}
I^{(2)} = 
  \int  d^d l_1\, d^d l_2
   \frac{1} {  l_1^2 (l_1-p_1)^2 (l_1-p_1-p_2)^2  l_2^2 (l_2+p_4)^2 (l_2+p_3+p_4)^2
             (l_1-l_2)^2}\,, 
\hskip 1 cm 
\label{DoubleBox}
\end{equation}
illustrated in \fig{fig:dbox-dual}.  We define the dual points implicitly, via
\begin{align}
& p_1 = x_2 - x_1\,, \hskip 1cm 
p_2 = x_3 - x_2\,, \hskip 1cm 
p_3 = x_4 - x_3\,, \hskip 1cm 
p_4 = x_1 - x_4\,, \nonumber \\
\hskip 1cm 
& l_1 = y_1 - x_1\,, \hskip 1cm
l_2 = y_2 - x_1\,.
\label{twoloopktox}
\end{align}
These variables automatically enforce momentum conservation on the $p_i$.
Performing the change of variables~(\ref{twoloopktox})
in the double box, gives,
\begin{equation}
I^{(2)} =  \int d^d y_1\, d^d y_2 \,
\frac{1}{(y_1 - x_1)^2 (y_1 - x_2)^2 (y_1 - x_3)^2 (y_1-y_2)^2 (y_2 - x_1)^2 (y_2 - x_3)^2 (y_2 -x_4)^2}\,.
\label{TwoLoopDualForm}
\end{equation}
The dual conformal transformations include scalings $z_i \rightarrow a
z_i$ and inversions $z_i^\mu \rightarrow {z_i^\mu / z_i^2}$, where $z_i$ may be
either an external $x_i$ or internal $y_A$ dual point.  Under the
inversion, we have
\begin{equation}
(x_{i}-x_{j})^2 \rightarrow {(x_{i}-x_{j})^2 \over x_i^2 x_j^2} \,, \hskip 1 cm
(y_{A}-x_{j})^2 \rightarrow {(y_{A}-x_{j})^2 \over y_A^2 x_j^2} \,, \hskip 1 cm
(y_{A}-y_{B})^2 \rightarrow {(y_{A}-y_{B})^2 \over y_A^2 y_B^2} \,.
\end{equation}
From the perspective of planar $\NeqFour$ super-Yang-Mills theory,
dual conformal transformations are interesting because they formally
leave the amplitude invariant, ignoring regulator issues.  From our
perspective, what makes them interesting is that they leave the
unitarity cut surface $(y_A - x_j)^2=0$ invariant.  These
considerations suggest that we can generate IBP relations and differential equations that are
automatically compatible with unitarity.  This is true whether or not
the integrals are invariant.  Indeed, the noninvariance is precisely
what we will use to generate nontrivial IBP
relations and differential equations.

To generate IBP relations and differential equations we should phrase the conformal
transformation as infinitesimal transformations.  Under an
infinitesimal conformal boost defined by an $\SO(d-1,1)$ vector $b^\mu$,
a dual coordinate $z^\mu$ 
transforms as
\begin{equation}
\Delta z^\mu = \frac 1 2 b^\mu z^2 - (b \cdot z) z^\mu \,. 
\label{eq:boostFormula}
\end{equation}
Under an infinitesimal scaling (i.e.\ dilatation) transformation with parameter $\beta$,
\begin{equation}
\Delta z^\mu = \beta z^\mu \, . \label{eq:scaleFormula}
\end{equation}
(Throughout this paper, $\Delta$ will be understood as a differential
operator or a symmetry generator, so the right hand side of the above equation is
not multiplied by an explicit infinitesimal parameter.)  Finally,
under Lorentz transformations parametrized by an antisymmetric
$\Omega^{\mu \nu}$,
\begin{equation}
\Delta z^\mu = \Omega^{\mu \rho} \eta_{\rho \nu} z^\nu = \Omega^{\mu}_{\ \nu} \, z^\nu \, ,
\end{equation}
where $ \eta_{\rho \nu}$ is the metric.
Combining the conformal boost, scaling, and Lorentz transformations, we have
\begin{equation}
\Delta z^\mu =  \frac 1 2 b^\mu z^2 + (\beta - b \cdot z) z^\mu + \Omega^\mu_{\ \nu} \, z^\nu \, . \label{eq:totalTransform}
\end{equation}
In terms of the infinitesimal transformations, if two points $z_1^\mu$
and $z_2^\mu$ both transform according to
Eq.~\eqref{eq:totalTransform}, then a simple calculation gives
\begin{equation}
\Delta (z_1 - z_2)^2 = \left[ 2 \beta - b \cdot (z_1 + z_2) \right] (z_1 - z_2)^2 \,,
\label{eq:DualConfOnInvProp}
\end{equation}
which is proportional to $(z_1 - z_2)^2$. Therefore, under an
infinitesimal dual conformal transformation for $y_A^\mu$ and $x_j^\mu$,
the variation of any inverse propagator is proportional to the inverse
propagator itself. This immediately echoes the condition
Eq.~\eqref{eq:tangentDE} for the lack of propagators raised to higher powers, and
implies that dual conformal transformations generate
unitarity-compatible differential equations~\cite{DEnoDouble}. The dual-spacetime integration measure transforms as the trace of the infinitesimal deviation of the Jacobian matrix from the identity matrix,
\begin{equation}
\Delta \left( d^d z \right) = d^d z \frac{\partial \Delta z^\mu}{\partial z^\mu} 
  = d^d z \,\left( \beta - b \cdot z \right)d \,.
\label{eq:weightOfDz}
\end{equation}
As discussed in Subsection \ref{subsec:UCDE}, IBP-generating vectors arise if we impose the further condition that the infinitesimal dual conformal
transformations do not shift the external points,
\begin{align}
\Delta x_j &= 0\,,  
\label{eq:deltaExtZero}
\end{align}
for each external point $x_j$.  We will give examples in subsequent sections
for explicitly solving this constraint.

\subsection{Embedding formalism}
\label{EmbeddingSubsection}

A convenient means for carrying out conformal transformations is via
the embedding formalism of Refs.~\cite{SimmonsDuffin,
CaronHuotEmbedding}.  In this construction, the system is embedded in
a space with two extra dimensions.  This allows us to reformulate dual
conformal transformations as Lorentz transformations in the
higher-dimensional space.

The embedding formalism maps each dual point $z^\mu$ in the
$\SO(d-1,1)$ dual space to a point in $\SO(d,2)$ invariant space. Following the
conventions of Ref.~\cite{CaronHuotEmbedding}, we introduce
\begin{equation}
Z^a = \begin{pmatrix}
Z^\mu \\
Z^- \\
Z^+
\end{pmatrix}
= \begin{pmatrix}
z^\mu \\
-z^2 \\
1 \\
\end{pmatrix}. \label{eq:embed}
\end{equation}
These vectors are defined modulo the identification
\begin{equation}
Z \cong \alpha Z, \quad \alpha \neq 0 \,,
 \label{eq:gl1}
\end{equation}
which is referred to as a $\GL(1)$ ``gauge freedom''. The inverse map is
\begin{equation}
z^\mu = \frac{Z^\mu}{Z^+} \, . \label{eq:embedInverse}
\end{equation}
The $\SO(d,2)$ invariant contraction is defined by the inner product
\begin{equation}
(X Y) = X^a X_a \equiv 2 X^\mu Y_\mu + X^+ Y^- + X^- Y^+ \, .
\end{equation}
Thus the point defined in Eq.~\eqref{eq:embed} is on the lightcone,
\begin{equation}
(ZZ)=0 \, .
\end{equation}
We introduce the point at infinity, $I$, which is the limit of
Eq.~\eqref{eq:embed} with all components of $x^\mu$ uniformly tending
to infinity, with an appropriate scaling using the gauge freedom in
Eq.~\eqref{eq:gl1},
\begin{equation}
I^a = \lim_{|z| \to \infty} \left( -\frac{1}{z^2} \right)
\begin{pmatrix}
z^\mu \\
-z^2 \\
1
\end{pmatrix}
=
\begin{pmatrix}
0 \\
1 \\
0
\end{pmatrix}  .\label{eq:inf}
\end{equation}
This has the effect of compactifying the loop-momentum
space~\cite{AbreuBrittoEtAl}.  Using \eqn{eq:embed}, we map the
loop-momentum points $y_A^\mu$ to
\begin{equation}
Y_A^a = \begin{pmatrix}
y_A^\mu \\
-y_A^2 \\
1
\end{pmatrix}, \label{eq:lmap}
\end{equation}
and map the dual kinematic points $x_j^\mu$ to
\begin{equation}
X_j^a = \begin{pmatrix}
x_j^\mu \\
-x_j^2 \\
1
\end{pmatrix} . \label{eq:qmap}
\end{equation}
The inverse propagators are now represented by $\SO(d,2)$ inner products between these points,
\begin{align}
(y_A-y_B)^2 &= -\frac{(Y_A Y_B) }{(Y_A I) (Y_B I)}\,, \label{eq:llmap}\\
(y_A - x_j)^2 &= -\frac{(Y_A X_j) }{(Y_A I) (X_j I)}\,, \label{eq:lqmap}
\end{align}
where $\GL(1)$ invariance is ensured by the denominators involving
the point at infinity. The denominators are unity in the gauge of
Eq.~\eqref{eq:lmap}. The factor $(X_j I)$ in the
denominator of the right hand side of the second line can be omitted, because we will always
choose the gauge $(X_j I) =1$, as in \eqn{eq:qmap}.

The integration measure for each loop becomes, suppressing the loop label,
\begin{equation}
d^d y \rightarrow \frac{ d^{d+2} Y \, \delta (Y^2 /2 )}{(YI)^d \operatorname{Vol}(\GL(1))} \, ,
 \label{eq:measure}
\end{equation}
where $Y^2$ is a shorthand for $(Y Y) = Y^a Y_a$ and 
the expression is formally divided by the volume of the $\GL(1)$ gauge orbit. 

We define $\SO(d,2)$ Lorentz transformations acting on some function
$f(Z)$ using two reference vectors $Z_i$ and $Z_j$, as
\begin{align}
\Delta f(Z) &= (Z_{[i}\, Z) \, \left(Z_{j]} \, \partialdisplay Z \right) f(Z) 
= Z^a_{[i}\, Z_a \, Z^b_{j]} \, \partialdisplay {Z^b} f(Z) \nn \\
&=  \biggl(Z^a_{i}\, Z_a \, Z^b_{j} \, \partialdisplay {Z^b} - Z^a_{j}\, Z_a \, Z^b_{i}  \, \partialdisplay {Z^b}
\biggr) f(Z) \,,
 \label{eq:deltaZ}
\end{align}
where $a$ and $b$ are $\SO(d,2)$ indices.
Notice that the factor $\delta(Y^2 /2)$ in Eq.~\eqref{eq:measure} is
invariant under these transformations.  The square-bracket notation in
the first line indicates antisymmetrization over $i$ and $j$, as
explicitly implemented in the second line.

Integration-by-parts relations follow from Lorentz invariance identities~\cite{CaronHuotEmbedding},
\begin{equation}
0= \int \frac{ d^{d+2} Y \, \delta (Y^2 /2 )}{\operatorname{Vol}(\GL(1))} \, u(Z_i, Z_j) \mathcal I \,, 
\label{eq:embedLI}
\end{equation}
where 
\begin{equation}
u(Z_i, Z_j)\equiv  (Z_{[i}\, Y) \, \biggl(Z_{j]} \, \partialdisplay Y \biggr) 
= (Z_i^b Y_b Z_j^a - Z_j^b Y_b Z_i^a) \partialdisplay{Y^a} \,,
\label{eq:uxixj}
\end{equation}
is a one-loop IBP-generating vector. In \eqn{eq:embedLI} it acts on some
general loop integrand $\mathcal I$. The factor $1/(YI)^d$ from the
integration measure in Eq.~\eqref{eq:measure} is absorbed into
$\mathcal I$. Concrete examples of such IBP relations will be given in
subsequent sections.

The $\SO(d,2)$ Lorentz transformations exactly
correspond to conformal transformations in Minkowski space with
$\SO(d-1,1)$ invariant metric, which can be checked using the inverse
map formula Eq.~\eqref{eq:embedInverse}. For example, in
Eq.~\eqref{eq:uxixj}, a $d$-dimensional translation $\Delta z^\mu =
e^\mu$ is equivalent to setting
\begin{equation}
Z_i= I = \begin{pmatrix} 0\\ 1 \\ 0 \end{pmatrix}, \hskip 1 cm Z_j= \begin{pmatrix} e^\mu \\ 0\\ 0 \end{pmatrix} .
\end{equation}
A $d$-dimensional conformal boost Eq.~\eqref{eq:boostFormula} with parameter $b^\mu$ is equivalent to setting
\begin{equation}
Z_i = -\frac 1 2 \begin{pmatrix} 0 \\ 0 \\ 1 \end{pmatrix}, \hskip 1 cm Z_j = \begin{pmatrix} b^\mu \\ 0 \\ 0 \end{pmatrix} .
\end{equation}
Finally, a scaling transformation Eq.~\eqref{eq:scaleFormula} is equivalent to setting
\begin{equation}
Z_i = I = \begin{pmatrix} 0\\ 1 \\ 0 \end{pmatrix}, \hskip 1 cm Z_j = -\begin{pmatrix} 0\\ 0 \\ \beta \end{pmatrix} .
\end{equation}
Therefore the IBP relations from $\SO(d,2)$ Lorentz invariance arise
from infinitesimal conformal transformations of the $d$-dimensional
loop momenta. Following the logic of the previous subsection, such IBP
relations will not have propagators raised to higher powers if the $\SO(d,2)$ Lorentz
transformations in Eq.~\eqref{eq:deltaZ} leaves the external momenta
invariant, i.e.\ leaves the $X_j$ points invariant up to $GL(1)$ gauge
scaling.

More generally, we can consider any IBP-generating vector in the embedding space,
\begin{equation}
V^a \partialdisplay{Y^a} \, .
\end{equation}
The above expression can be identified with an IBP-generating vector $v^\mu \partial_\mu$ in ordinary $\SO(d-1,1)$ space if it satisfies the following two conditions: (i) it must be $\GL(1)$ gauge-invariant, and (ii) it must commute with the measure factor $\delta(Y^2/2)$, i.e.,
\begin{equation}
V^a Y_a = 0 \,.
\end{equation}
The resulting IBP relation is, again showing the one-loop case for illustration,
\begin{align}
0 &= \int \frac{ d^{d+2} Y \, \delta (Y^2 /2 )}{\operatorname{Vol}(\GL(1))} \partialdisplay{Y^a} 
\left( V^a \mathcal I \right) \nn \\
&= \int \frac{ d^{d+2} Y \, \delta (Y^2 /2 )}{\operatorname{Vol}(\GL(1))} \left( \mathcal I \, \frac {\partial V^a}{\partial Y^a} + V^a \frac{\partial \mathcal I}{\partial Y^a} \right), 
\label{eq:ibpAsDivergence}
\end{align}
consisting of a divergence term proportional to an integrand $\mathcal I$ and a second
term involving derivatives of $\mathcal I$. For an IBP-generating
vector as in Eq.~\eqref{eq:uxixj} from Lorentz invariance, the divergence
term vanishes, so Eq.~\eqref{eq:embedLI} only involves derivatives of
$\mathcal I$.

We can extend the above discussion to include internal
masses~\cite{AbreuBrittoEtAl, AldayMassiveEmbedding} by modifying
Eq.~\eqref{eq:qmap} to map the external momentum point $x_j^\mu$ to
\begin{equation}
X_j^a = \begin{pmatrix}
x_j^\mu \\
-x_j^2 + m_j^2\\
1
\end{pmatrix} \, . \label{eq:qmapMass}
\end{equation}
This changes Eq.~\eqref{eq:lqmap} to
\begin{equation}
(x_j - y_A)^2 - m_j^2= -\frac{(Y_A X_j) }{(Y_A I)} \, . \label{eq:lqmapMass}
\end{equation}
Since Eq.~\eqref{eq:lqmapMass} contains a mass $m_j$ that is independent of the loop label $A$, the formula only allows arbitrary masses at the one-loop level, and at higher loops, the masses of some propagators must be correlated or vanishing.

\section{IBP for one-loop triangle integrals}
\label{sec:triangle}

To illustrate the ideas of the previous section, we present some simple one-loop examples. It is well known that by Passarino-Veltman or OPP reduction\cite{PV,OPP,Forde}, triangle tensor integrals can all be reduced to triangle scalar integrals and daughter integrals (i.e.\ bubble and tadpole integrals from collapsing certain propagators of the triangle diagram). In the language of unitarity-compatible IBP reduction, this is accomplished by IBP-generating vectors which are rotation generators in the spacetime directions orthogonal to all external momenta~\cite{ItaIBP}. However, under special kinematic configurations, scalar triangle integrals can be further reduced to bubble integrals using IBP. As will be shown in Section \ref{sec:Landau} near the end of the paper, these special kinematic configurations are exactly those which allow leading Landau singularities. The necessary IBP-generating vectors will be the main topic of this section.

First we show directly how dual conformal transformations can be used
to generate unitarity-compatible IBP relations without higher-power
propagators.  We then streamline the procedure using the embedding
formalism~\cite{SimmonsDuffin, CaronHuotEmbedding} that reduced
conformal transformations to simpler Lorentz transformations in higher
dimensions.

\subsection{One-external-mass triangle: direct treatment}
\label{subsec:direct}

\begin{figure}
  \centering
  \includegraphics[width=0.25\textwidth]{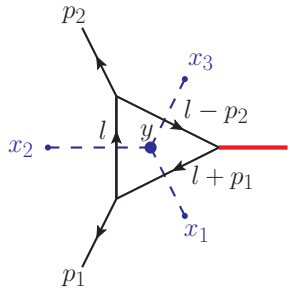}
  \caption{The one-loop triangle with outgoing external momenta $p_1,
    \, p_2, \,-p_1-p_2$ and dual points $x_1,x_2,x_3$. All internal propagators are massless, and
    the single massive external leg has mass $(p_1+p_2)^2=s$, shown as
    a thick (red) line. The dashed (blue) lines indicate the dual diagram.}
  \label{fig:triangle}
\end{figure}

Consider the one-loop triangle shown in \fig{fig:triangle}. For
illustrative purposes, we specialize to the simple case of all
internal and external legs being massless, with the exception of the
right-most leg of the figure.  The three inverse propagators are
\begin{equation}
\D_1=(l+p_1)^2, \hskip 1 cm \D_2= l^2, \hskip 1 cm  \D_3= (l-p_2)^2 \,. 
\label{eq:triangleProp0}
\end{equation}
The external kinematic invariants are
\begin{equation}
p_1^2=p_2^2=0, \hskip 1.5 cm (p_1+p_2)^2 = s \, . \label{eq:triangleInv0}
\end{equation}
We introduce dual coordinates $x_j, \, 1 \leq j \leq 3$ and $y$, such that
\begin{equation}
p_1 = x_2 - x_1, \hskip 1 cm  p_2 = x_3 - x_2, \hskip 1 cm l = y - x_2
\, . \label{eq:trianglePfromX}
\end{equation}
The external points $x_j$ completely fix the external momenta,
while $y$ is an internal point corresponding to shifted loop
momentum.  Since $p_j$ and $l$ are expressed as differences
between dual coordinates $x_j$ and $y$ in
Eq.~\eqref{eq:trianglePfromX}, we are free to apply the same
translation to all the dual coordinates.  We choose to fix the
translation ``gauge freedom'' by taking,
\begin{equation}
x_2 = 0 \,,
\end{equation}
so the explicit expressions for the dual coordinates are
\begin{equation}
x_1 = -p_1\,, \hskip 1 cm x_2 = 0\,, \hskip 1 cm x_3 = p_2\,, \hskip 1 cm  y = l \,. \label{eq:triangleExplicitX}
\end{equation}
With this gauge choice, in terms of these dual coordinates,
\eqns{eq:triangleProp0}{eq:triangleInv0} become
\begin{equation}
\D_1 = (y-x_1)^2\,, \hskip 1 cm \D_2 = y^2\,, \hskip 1 cm \D_3 = (y-x_3)^2\,,
\label{eq:triangleProp}
\end{equation}
and
\begin{equation}
x_1^2 = x_2^2 = x_3^2 = 0, \hskip 1 cm (x_2-x_1)^2 = (x_3 - x_2)^2=0, \hskip 1 cm 
(x_3 - x_1)^2 = s \, . \label{eq:triangleInv}
\end{equation}
Eqs.~\eqref{eq:triangleProp} and \eqref{eq:triangleInv} imply that
\begin{equation}
y \cdot x_1 = \frac 1 2 (\D_2 - \D_1), \hskip 1 cm y \cdot x_2 =0\,,
 \hskip 1 cm  y \cdot x_3 = \frac 1 2 (\D_2 - \D_3) \, . 
\label{eq:triangleLQ}
\end{equation}

As discussed in \sect{sec:key}, the key property of the dual conformal
transformations \eqref{eq:totalTransform} is that when acting on
inverse propagators, they return results proportional to the inverse
propagator itself, as shown in \eqn{eq:DualConfOnInvProp}.  In order
to use dual conformal transformations to generate IBP
relations, we restrict to the subset \eqref{eq:deltaExtZero}
where the transformations do not shift the external points.

The shift under the transformation of the loop momentum gives an
IBP-generating vector
\begin{align}
\Delta y^\mu \partialdisplay{y^\mu}\ &= \Delta l^\mu \partialdisplay{l^\mu} 
= v^\mu\partialdisplay{l^\mu} \,,
\end{align}
that satisfies the key condition of \eqn{eq:tangent} that 
it does not raise the power of propagators in the IBP identity.
Applying \eqn{eq:totalTransform} to $x_j^\mu$, and using $x_j^2 = 0$ from \eqn{eq:triangleInv}, \eqn{eq:deltaExtZero} becomes
\begin{equation}
0 = (\beta - b \cdot x_j) x_j, \quad j=1,2,3 \, . \label{eq:deltaQjZero1}
\end{equation}
One solution to Eq.\ \eqref{eq:deltaQjZero1} is
\begin{equation}
\beta = s\,, \hskip 2 cm 
b = -2 (x_1 + x_3) \,, \label{eq:triangleBsol}
\end{equation}
where we used \eqn{eq:vector}.
This gives, using Eq.\ \eqref{eq:totalTransform},
\begin{equation}
\Delta l = v = - l^2 ( x_1 + x_3 ) + \left[ s + 2 \, l \, \cdot (x_1 + x_3) \right] l \, .
\end{equation}
The IBP-generating vector $v^\mu \partial_\mu$ satisfies
\begin{equation}
v^\mu \partialdisplay{l^\mu} \rho_i = \mathcal W_i \rho_i, \quad 1\leq i \leq 3 \,,
\end{equation}
where $\mathcal W_i$ follows from Eq.\ \eqref{eq:DualConfOnInvProp},
\begin{equation}
\mathcal W_i = 2 \beta - b \cdot (l + x_i) \, .
\end{equation}
The divergence of the vector follows from Eq.\ \eqref{eq:weightOfDz},
\begin{equation}
\mathcal W_{\rm measure} = \frac{\partial v^\mu}{\partial l^\mu} = (\beta - b \cdot l) d \, .
\end{equation}
We obtain the IBP relation
\begin{align}
0 &= \int d^d l \,\partialdisplay{l^\mu} \frac{v^\mu}{\D_1 \D_2 \D_3} \nn \\
&= \int d^d l \, \left( \mathcal W_{\rm measure} - \mathcal W_1 - \mathcal W_2 - \mathcal W_3 \right) \frac{1}{\D_1 \D_2 \D_3} \nn \\
&= \int d^d l \, \big[ (d-6)\beta + b \cdot (x_1+x_2+x_3) - (d-3) b \cdot l \big] \frac{1}{\D_1 \D_2 \D_3} \nn \\
&= \int d^d l \, \big[ (d-4)s + 2(d-3)(x_1+x_3) \cdot l \big] \frac{1}{\D_1 \D_2 \D_3} \, .
\end{align}
In the last line above, we have used the explicit solution for $\beta$
and $b^\mu$ in Eq.\ \eqref{eq:triangleBsol}. Simplifying the result
using Eqs.~\eqref{eq:triangleExplicitX} and~\eqref{eq:triangleLQ}, the
final IBP relation is
\begin{align}
0 &= \int d^d l \big[ (d-4)s + (d-3)(2\rho_2 - \rho_1 - \rho_3) \big] \frac{1}{\D_1 \D_2 \D_3} \nn \\
&= (d-4) s I_{\rm tri} + 2 (d-3) I_{\rm bub}^{(s)} \,, \label{eq:ibpTri}
\end{align}
where $I_{\rm tri}$ is the scalar triangle integral
in \fig{fig:triangle} and $I_{\rm bub}^{(s)}$ is the scalar bubble
integral obtained from the term proportional to $\D_2$ which cancels
the propagator $1/\D_2 = 1/l^2$, so that the mass of both external
legs is $s$. The terms proportional to $\D_1$ and $\D_3$ in the second
line of Eq.\ \eqref{eq:ibpTri} give bubble integrals with massless
external legs, which vanish in dimensional regularization and are
discarded in the last line.  \Eqn{eq:ibpTri} corresponds to a well
known relation between the one-external-mass triangle and the bubble
integral (see e.g.\ the fourth appendix of Ref.~\cite{BernChalmers}). The coefficient of the triangle integral in \Eqn{eq:ibpTri} vanishes as $d \to 4$ while the coefficient of the bubble integral does not. This is due to infrared singularities of the triangle integral. In fact, the triangle integral allows a leading Landau singularity, whose connection with IBP-generating vectors will be explored in Section \ref{sec:Landau}.
This simple example illustrates the basic principle behind using dual
conformal symmetry to generate useful IBP relations.

\subsection{Embedding-space treatment of one- and two-external-mass triangle}
\label{subsec:embedTri}

To streamline dual conformal transformations and the construction of
IBP-generating vectors we use the embedding
formalism~\cite{SimmonsDuffin, CaronHuotEmbedding} summarized in
\sect{EmbeddingSubsection}.  This reduces $d$-dimensional conformal
transformations to simpler $(d+2)$-dimensional Lorentz
transformations.  The algorithm involves solving for all
$(d+2)$-dimensional Lorentz transformations that leave the external
momenta invariant. This is used to construct a matrix that encodes the
action of the IBP vector, so that the IBP relations can be
conveniently constructed. We will use the above one-external-mass triangle as
an example, before explaining the generalization.  Here we apply
Lorentz rotations that act in the subspace of external points.  One
can also consider Lorentz rotations in the embedding space that only
act in the space orthogonal to the external points, as we do in
\sect{subsec:generalities}.

Using Eqs.~\eqref{eq:lqmap} and~\eqref{eq:measure}, the scalar
triangle integral in \fig{fig:triangle} is written in the
$\SO(d,2)$ embedding space,
\begin{equation}
I_{\rm tri}= \int \frac{d^{d+2} Y \, \delta (Y^2 /2 )}{(YI)^{d-3} \operatorname{Vol}(\GL(1))} 
\, \frac{(-1)^3}{(YX_1)(YX_2)(YX_3)}\,, 
\label{eq:TriEmbed}
\end{equation}
where $Y$ and $X_j$ are as defined in Eqs.\ \eqref{eq:lmap} and \eqref{eq:qmap} and
as in \eqn{eq:measure} $Y^2$ is a shorthand for $(Y Y) = Y^a Y_a$. 
The factor $(-1)^3$ comes from the minus sign on the right hand side of Eq.\ \eqref{eq:lqmap}.

We define a subset of infinitesimal $d+2$ dimensional Lorentz
transformations $\Delta_\omega$ by an antisymmetric $4 \times 4$ matrix
$\omega^{ij}$, acting on a $(d+2)$ dimensional point $Z^a$ as
\begin{equation}
\Delta_\omega Z^a = \sum_{1\leq i,j \leq 4} (Z X_i) \omega^{ij} X_j^a \, , \label{eq:ibpAnsatz}
\end{equation}
where $X_1, X_2, X_3$ are the three external points in
\eqn{eq:TriEmbed} and $X_4 = I$, where $I$ is defined in \eqn{eq:inf}.
We will choose the $\omega^{ij}$ such that the above Lorentz
transformation leaves $X_1, X_2, X_3$ invariant up to a GL$(1)$ gauge
scaling. This means that under the transformations only the
loop-momentum shifts by an infinitesimal amount, captured by the IBP
vector,
\begin{equation}
\frac 1 2 \omega^{ij} u(X_i, X_j) = \omega^{ij} (X_i Y) X_j^a \partialdisplay{Y^a} \,.
\label{eq:omegaVec}
\end{equation}
The summation over $1\leq i,j\leq 4$ is implicit, and we have
used the definition of $u(X_i, X_j)$ in Eq.\ \eqref{eq:uxixj}.

The Lorentz transformations in Eq.\ \eqref{eq:ibpAnsatz} acts on $X_k$ as
\begin{align}
\Delta_\omega X_k^a &= g_{ki} \, \omega^{ij}  X_j^a  \nonumber \\
&\equiv \bar \omega_k^{\ j} X_j^a\,,
 \label{eq:ibpAction}
\end{align}
where we defined the ``embedding space gram matrix'' as, 
\begin{equation}
g_{ij} = (X_i X_j) =
 \begin{pmatrix}
  0 & 0 & -s & 1 \\
  0 & 0 & 0 & 1 \\
  -s & 0 & 0 & 1 \\
  1 & 1 & 1 & 0
 \end{pmatrix}, \label{eq:dualGramMat}
\end{equation}
where we identify $X_4$ with $I$ and the last row and column contain
entries of unity due to the gauge choice $(X_j I)=1$ in
Eq.\ \eqref{eq:qmap}.  We then impose the condition that $X_1, X_2,
X_3$ but not $X_4 = I$, are left invariant by the Lorentz
transformation:
\begin{equation}
\Delta_\omega X_k^a = \alpha_k \, X_k^a \,, \hskip 1 cm \text{if } k=1,2,3,
 \label{eq:invUpToGL1}
\end{equation}
where $\alpha_k$ can be absorbed into the $GL(1)$ invariance of the
integrand \eqref{eq:gl1} which takes $(d+2)$-dimensional vectors to be
equivalent if they are scaled.  The second line in \eqn{eq:ibpAction}
defines the ``IBP matrix'', and depends on the free parameters
$\omega^{ij}$ which we determine below,
\begin{equation}
\bar \omega \equiv g \, \omega = \begin{pmatrix}
   s \omega_{13}-\omega_{14} & s \omega_{23}-\omega_{24} & -\omega_{34} & -s \omega_{34} \\
  -\omega_{14} & -\omega_{24} & -\omega_{34} & 0 \\
  -\omega_{14} & -s \omega_{12}-\omega_{24} & -s \omega_{13}-\omega_{34} & -s \omega_{14} \\
  -\omega_{12}-\omega_{13} & \omega_{12}-\omega_{23} & \omega_{13}+\omega_{23} & \omega_{14}+\omega_{24}+\omega_{34}
  \end{pmatrix} .
\label{eq:barOmegaDef}
\end{equation}
Eq.~\eqref{eq:invUpToGL1} implies
\begin{equation}
\bar \omega_k^{\ j} = 0 \quad \text{if } k=1,2,3, \, j \neq k, 1 \leq j \leq 4 \,, 
\label{eq:barOmegaZeroCondition}
\end{equation}
i.e., the non-diagonal entries have to vanish in all but the last
rows.  This gives four independent homogeneous linear constraints on
the six possible components of the antisymmetric matrix $\omega$,
\begin{align}
\omega_{14} =\omega_{34}&=0 \,, \nn \\
-s \omega_{12} - \omega_{24} &= 0\,, \nn \\
s \omega_{23} - \omega_{24} &= 0 \, .
\end{align}
The two independent solutions are
\begin{equation}
\omega^{(1)} = \begin{pmatrix}
   0 & -1 & 0 & 0 \\
   1 & 0 & 1 & s \\
   0 & -1 & 0 & 0 \\
  0 & -s & 0 & 0
 \end{pmatrix}, \hskip 1 cm 
\omega^{(2)} = \begin{pmatrix}
   0 & 0 & 1 & 0 \\
   0 & 0 & 0 & 0 \\
   -1 & 0 & 0 & 0 \\
   0 & 0 & 0 & 0
 \end{pmatrix}, \label{eq:triangleOmega}
\end{equation}
under which the IBP matrix in Eq.\ \eqref{eq:barOmegaDef} becomes
\begin{equation}
\bar \omega^{(1)} = \begin{pmatrix}
  0 & 0 & 0 & 0 \\
  0 & -s & 0 & 0 \\
  0 & 0 & 0 & 0 \\
  1 & -2 & 1 & s
 \end{pmatrix}, \hskip 1 cm 
\bar \omega^{(2)} = \begin{pmatrix}
  s & 0 & 0 & 0 \\
  0 & 0 & 0 & 0 \\
  0 & 0 & -s & 0 \\
  -1 & 0 & 1 & 0
 \end{pmatrix}, \label{eq:triangleOmegaBar}
\end{equation}
respectively.
To compute IBP relations, the IBP vector Eq.\ \eqref{eq:omegaVec} acts on $(Y X_k)$ as
\begin{align}
\Delta_\omega  (Y X_k) &= \frac 1 2 \omega^{ij} u(X_i, X_j) (Y X_k) = \omega^{ij}(X_i Y) (X_j X_k) = -g_{kj} \omega^{ji} (YX_i)\nn \\
&= - \bar \omega_k^{\ i} (Y X_i) , \label{eq:omegaVecOnYX}
\end{align}
where we used the antisymmetry of $\omega$, and $u(X_i, X_j)$ is
defined in \eqn{eq:uxixj}.  In terms of matrix components $\bar
\omega_i{}^j$ that are nonvanishing for either solution, the resulting
IBP relation is,
\begin{align}
  0&=\int \frac{d^{d+2} Y \, \delta (Y^2 /2 )}{\operatorname{Vol}(\GL(1))} \, 
    \Delta_{\omega} \biggl(\frac{(-1)^3}{(YI)^{d-3} (YX_1)(YX_2)(YX_3)} \biggr) \nn \\
&= \int \frac{d^{d+2} Y \, \delta (Y^2 /2 )}{\operatorname{Vol}(\GL(1))} \,  \frac{(-1)^3}
  {(YI)^{d-3} (YX_1)(YX_2)(YX_3)}  
\Biggl\{ \biggl[ \biggl( \sum_{i=1}^3 \bar \omega_i^{\ i} \biggr) + (d-3) \bar \omega_4^{\ 4} \biggr] \nn \\
& \hskip 4.5 cm \null
  +\frac{1}{(YI)} \biggl[ (d-3) \sum_{i=1}^3 \bar \omega_4^{\ i} (Y X_i) \biggr] \Biggr\}
 \nn \\
&= \int d^d l \,\frac{1}{\D_1 \D_2 \D_3} \Biggl\{
\biggl[ \biggl( \sum_{i=1}^3 \bar \omega_i^{\ i} \biggr) + (d-3) \bar \omega_4^{\ 4} \biggr] 
-(d-3) (\bar \omega_4^{\ 1} \D_1 + \bar \omega_4^{\ 2} \D_2  + \bar \omega_4^{\ 3} \D_3 )\Biggr\}\nn \\
&= \int d^d l \, \frac{1}{\D_1 \D_2 \D_3} \Biggl\{
\biggl[ \biggl( \sum_{i=1}^3 \bar \omega_i^{\ i} \biggr) + (d-3) \bar \omega_4^{\ 4} \biggr] 
- (d-3) \bar \omega_4^{\ 2} \D_2  \Biggr\} \,,
\label{IBPTri1m}
\end{align}
where on the last line we dropped the contributions proportional to $\D_1$ and $\D_3$ 
because those generate scaleless bubble integrals that vanish in dimensional regularization.

Substituting the first solution for $\bar\omega$ in \eqn{eq:triangleOmegaBar} , 
\begin{equation}
 {\bar \omega}^{(1)}_{\;1}{}^1 = {\bar \omega}^{(1)}_{\; 3}{}^3 = 0\,,
   \hskip 1.5 cm  {\bar \omega}^{(1)}_{\; 4}{}^4 = - {\bar \omega}^{(1)}_{\; 2}{}^2 = s, 
\hskip 1.5 cm {\bar \omega}^{(1)}_{\; 4}{}^{2} = -2 \,,
\end{equation}
into \eqn{IBPTri1m} yields,
\begin{equation}
0 = s(d-4) I_{\rm tri} + 2(d-3) I_{\rm bub}^{(s)}\,, 
\label{eq:ibpTriAgain}
\end{equation}
reproducing Eq.~\eqref{eq:ibpTri}.

\begin{figure}
  \centering
  \includegraphics[width=0.3\textwidth]{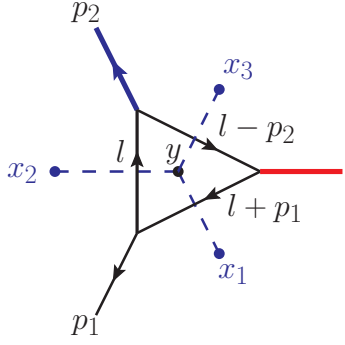}
  \caption{The one-loop triangle with outgoing external momenta $p_1,
    \, p_2, \, -p_1-p_2$. All internal propagators are massless, and
    the massive external legs, shown as thick lines, have masses
    $p_2^2 =t$ and $(-p_1-p_2)^2=s$.  The dashed (blue) line indicates
    the dual diagram.}
  \label{fig:triangleTwo}
\end{figure}

For the second solution in Eq.\ \eqref{eq:triangleOmega}, we have 
\begin{equation}
 {\bar \omega}^{(2)}_{\;1}{}^1 = - {\bar \omega}^{(2)}_{\; 3}{}^3 = s\,,
  \hskip 1.5 cm  {\bar \omega}^{(1)}_{\; 2}{}^2 ={\bar \omega}^{(1)}_{\; 4}{}^4 = 0\,, \hskip 1.5 cm
 \hskip 1.5 cm {\bar \omega}^{(1)}_{\; 4}{}^{2} = 0  \,,
\end{equation}
so the IBP relation \eqref{IBPTri1m} is trivial because it involves
only integrals that vanish in dimensional regularization.

As another example, consider the triangle with two external
masses shown in \fig{fig:triangleTwo}. Following the same procedure as in the
one-external-mass case, we introduce dual coordinates as usual
\begin{equation}
p_1= x_2 - x_1\,, \hskip 1.5 cm p_2 = x_3 -x_2\,, \hskip 1.5 cm l = y - x_2\,.
\end{equation}
Following a similar analysis as for the single-external-mass case, we find
only a single solution that leaves all the external momenta invariant.
The associated IBP matrix is
\begin{equation}
\bar \omega = \begin{pmatrix}
-(s-t) & 0 & 0 & 0 \\
0 & -(s-t) & 0 & 0 \\
0 & 0 & s-t & 0 \\
2 & -2 & 0 & s-t
\end{pmatrix} ,
\end{equation}
which encodes the action of the IBP-generating vector through \eqn{eq:omegaVecOnYX}.
The resulting IBP relation, expressed in terms of the non-vanishing matrix components $\bar \omega_i^{\ j}$, is
\begin{align}
0&= \int d^d l \frac{1}{\D_1 \D_2 \D_3} \biggl\{
\biggl[  \sum_{i=1}^3 \bar \omega_i^{\ i}  + (d-3) \bar \omega_4^{\ 4} \biggr] 
-(d-3) \bar \omega_4^{\ 1} \D_1 - (d-3) \bar \omega_4^{\ 2} \D_2  - (d-3) \bar \omega_4^{\ 3} \D_3 \biggr\} \nn \\
&= (d-4) (s-t) I_{\rm tri}^{(s,t)} -2(d-3) I_{\rm bub}^{(t)} + 2(d-3) I_{\rm bub}^{(s)} \,, 
\label{eq:ibpTriMassive}
\end{align}
where $I_{\rm bub}^{(s)}$ is the bubble diagram obtained by canceling
the propagator $l-q_2$, and $I_{\rm bub}^{(s)}$ is the bubble diagram
obtained by canceling the propagator $l-q_1$. When $t=0$,
$I^{(t)}_{\rm bub}$ is a scaleless integral which vanishes in
dimensional regularization, so the above IBP relation
Eq.\ \eqref{eq:ibpTriMassive} becomes the same as the previous IBP
relation Eq.\ \eqref{eq:ibpTriAgain} found for the triangle with only
one massive external leg.

\subsection{The Higgs to $b \bar b$ decay triangle}
\label{subsec:Hbb}

\begin{figure}
  \centering
  \includegraphics[width=0.3\textwidth]{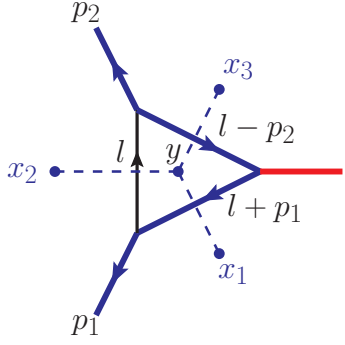}
  \caption{The one-loop triangle that appears in the decay of a Higgs
    boson to a $b \bar b$ quark pair. The outgoing external momenta
    are $p_1, \, p_2, \, -p_1-p_2$. The Higgs leg, shown as a thick
    (red) line on the rightmost part of the figure, has squared mass
    $(-p_1-p_2)^2 = m_H^2$. The bottom-quark lines, appearing in both
    external legs and internal propagators, are shown as thick (blue)
    lines with squared mass $m_b^2$. 
      }
  \label{fig:Hbb}
\end{figure}

As a more sophisticated example to illustrate the use of dual
conformal transformations in the presence of a mass, consider the
one-loop triangle integral involved in the decay of the Higgs to a $b
\bar b$ quark pair, with the bottom quark mass appearing in both
internal and external lines, as depicted in Fig.\ \ref{fig:Hbb}.
Internal masses are included in the embedding formalism, as described
at the end of \sect{EmbeddingSubsection}.

Introducing dual coordinates as usual, the three propagators are
written as squared differences between dual coordinates,
\begin{equation}
\D_1=(y-x_1)^2, \hskip 2 cm \D_2= (y-x_2)^2, \hskip 2 cm \D_3= (y-x_3)^2 \,, \label{eq:HbbProp}
\end{equation}
with gauge choice $x_2 = 0$, 
while the kinematic invariants are written as
\begin{equation}
x_2^2=0, \hskip 1 cm x_1^2 = x_3^2= m_b^2, \hskip 1 cm  (x_2-x_1)^2 = (x_3-x_2)^2 = 0, \hskip 1 cm 
  (x_1-x_3)^2 = m_H^2 \,, \label{eq:HbbInv}
\end{equation}
involving three massive external legs and two massive internal propagators.
Eqs.\ \eqref{eq:HbbProp} and \eqref{eq:HbbInv} imply that
\begin{equation}
y \cdot x_1 = \frac 1 2 (\D_2 - \D_1 +m_b^2)\,,
\hskip 1 cm y \cdot x_2 = 0\, , 
\hskip 1 cm y \cdot x_3 = \frac 1 2 (\D_2 - \D_3 + m_b^2) \,. 
\label{eq:HbbLQ}
\end{equation}
The embedding space Gram matrix is, using the mapping Eq.\ \eqref{eq:qmapMass} for the massive case and identifying $I$ with $X_4$,
\begin{equation}
g_{ij} = (X_i X_j) =
 \begin{pmatrix}
  2m_b^2 & 0 & 2m_b^2 - m_H^2 & 1 \\
  0 & 0 & 0 & 1 \\
  2m_b^2-m_H^2 & 0 & 2m_b^2 & 1 \\
  1 & 1 & 1 & 0
 \end{pmatrix}. \label{eq:dualGramMatHbb}
\end{equation}
Using the general algorithm illustrated in Subsection
\ref{subsec:embedTri}, there is only one solution to the antisymmetric
matrix $\omega^{ij}$ such that the IBP vector
\begin{equation}
\frac 1 2 \omega^{ij} u(X_i, X_j)\,,
\end{equation}
leaves all external momenta invariant. The solution is
\begin{equation}
\omega = \begin{pmatrix}
 0 & -1 & 0 & 0 \\
1 & 0 & 1 & m_H^2-4m_b^2 \\
0 & -1 & 0 & 0 \\
0 & -(m_H^2-4m_b^2) & 0 & 0
\end{pmatrix},
\end{equation}
which gives the IBP matrix,
\begin{equation}
\bar \omega = \begin{pmatrix}
0 & 0 & 0 & 0 \\
0 & -(m_H^2-4m_b^2) & 0 & 0 \\
0 & 0 & 0 & 0 \\
1 & -2 & 1 & (m_H^2 - 4m_b^2)
\end{pmatrix},
\end{equation}
which encodes the action of the IBP-generating vector through Eq.\ \eqref{eq:omegaVecOnYX}. The resulting IBP relation, expressed in terms of the matrix components $\bar \omega_i^{\ j}$, is
\begin{align}
0&= \int \! d^d l \, \frac{1}{\D_1 \D_2 \D_3} \left\{
\biggl[ \biggl( \sum_{i=1}^3 \bar \omega_i^{\ i} \biggr) + (d-3) \bar \omega_4^{\ 4} \biggr] 
-(d-3) \bar \omega_4^{\ 1} \D_1 - (d-3) \bar \omega_4^{\ 2} \D_2  - (d-3) \bar \omega_4^{\ 3} \D_3 \right \} \nn \\
&= (d-4) (m_H^2-4m_b^2) I_{\rm tri}^{Hb\bar b} +2(d-3) I_{\rm bub}^{(H)} - 2(d-3) I_{\rm bub}^{(b)}\,, \label{eq:ibpHbb}
\end{align}
where $I_{\rm tri}^{Hb\bar b}$ is the scalar triangle diagram, $I_{\rm
  bub}^{(H)}$ is the bubble sub-diagram obtained by canceling the
propagator with momentum $l$, and $I_{\rm bub}^{(b)}$ is the bubble
sub-diagram obtained by canceling either the propagator with momentum
$l+p_1$ or the one with momentum $l-p_2$. The IBP relation for the
one-external-mass triangle, Eq.\ \eqref{eq:ibpTriAgain} can be
reproduced from the above IBP relation Eq.\ \eqref{eq:ibpHbb} by
setting $m_H^2=s, \, m_b^2=0$.

For higher-loop planar integrals with up to four external legs of any
topology, the algorithm presented above can be adapted to find
nontrivial dual conformal transformations that leaves the all external
momenta invariant.  We start with the embedding-space Gram matrix for
the specific integral topology as in Eq.\ \eqref{eq:dualGramMat}, with
$X_{N+1}$ defined to be equal to $I$. Then we repeat the subsequent
calculations to produce the IBP matrix as in
Eq.\ \eqref{eq:barOmegaDef}, leading to homogeneous linear constraints
as in Eq.\ \eqref{eq:barOmegaZeroCondition}. Solving the linear
constraints gives the IBP vectors and relations. As discussed in the
previous subsection, for any solution of the antisymmetric matrix
$\omega$ that has a vanishing last column, we will obtain IBP
relations that only involve integrals with canceled propagators,
therefore such solutions may be discarded if we are interested in the
IBP reduction of top-level integrals.  In the next section we will
describe another class of useful dual conformal transformations
orthogonal to all external momenta, which will be useful at higher
loops.

\section{IBP for planar two-loop integrals}
\label{sec:planarTwoLoop}

In this section we discuss the more interesting case of higher-loop
integrals. With generic mass configurations (e.g.\ with all external and internal masses being different from each other), a complete set of IBP-generating vectors is tabulated in Ref.~\cite{ItaIBP}.  Here we apply dual conformal symmetry to uncover extra IBP-generating vectors for planar two-loop integrals with massless lines.
 In \sect{sec:nonplanar} we will extend this to the nonplanar case.

\subsection{Conformal transformations in transverse dimensions}
\label{subsec:boostTrans}

In the direct treatment of Section \ref{subsec:direct}, the parameter
of the conformal boost in Eq.\ \eqref{eq:triangleBsol}, with $x_i^\mu$
given in Eq.\ \eqref{eq:triangleExplicitX}, is a linear combination of
external momenta. However, another interesting possibility is
a conformal boost in a direction orthogonal to all external momenta,
which gives unitarity-compatible IBP-generating vectors for {\it
  every} planar integral at any loop order.

Consider a general $L$-loop $N$-point diagram. For a planar $N$-point
diagram at $L$ loops, we take the dual coordinates be
$x_1,x_2,\dots,x_N$. It is easy to fix the translation gauge freedom
such that every $x_i^\mu$ is written as a linear combination of the
external momenta $p_i^\mu$. (For example, if we fix $x_1 =0$,
then $x_i = \sum_{j=1}^{i-1} p_j$.) In
Eq.\ \eqref{eq:DualConfOnInvProp}, we choose the conformal-boost
parameter $b^\mu$ to be any vector that is orthogonal to all external
momenta, and do not include a scaling transformation (i.e.\ setting
$\beta=0$). This gives $\Delta (x_i - x_j)^2 =0$ for all pairs of
$i,j$, which means all Mandelstam variables are left
invariant. Therefore it is always possible to keep each individual
external momentum invariant by adding a compensating Lorentz
transformation.

In the planar case, the $\SO(d,2)$ embedding formalism gives a convenient
way of proceeding.  This eliminates the need to fix a gauge for the
translation degrees of freedom of the dual coordinates. For
illustration, we focus on $d=4-2\epsilon$ dimensional loop integrals with
$N$ external momenta, where $N \leq 5$. Generally, the embedding-space
reference points $X_1, X_2, \dots, X_N$ and the point at infinity $I$
together span $N+1$ ``physical'' dimensions, leaving an orthogonal
``transverse'' space of dimension $(d+2)-(N+1)=d-(N-1)$. This directly
corresponds to the subspace of ordinary $\SO(d-1,1)$ spacetime
orthogonal to the $N-1$ dimensions spanned by the external
momenta.\footnote{For example, for a five-point diagram, with
  dimensional regularization the transverse space has dimension
  $4-2\epsilon- (5-1) = -2\epsilon$.}  In addition, in the
$(N+1)$-dimensional ``physical'' space spanned by $X_1,X_2,\dots, X_N,
I$, one can always find one vector $\tilde I$ that satisfies the $N$
conditions,
\begin{equation}
(\tilde I X_i) = 0 , \quad 1 \leq i \leq N \, . \label{eq:tildeIconditions}
\end{equation}
In particular, if the top-left $N\times N$ sub-block $\tilde
g_{ij}=(X_i X_j)$ of the embedding-space Gram matrix is non-singular,
then the above $\tilde I$ can be found by projecting $I$ onto the
space orthogonal $X_1,X_2, \dots X_N$,
\begin{equation}
\tilde I^a = I^a - (I X_i) (\tilde g^{-1})_{ij} X_j^a \, . \label{eq:projectIperpXi}
\end{equation}

We can also define a set of vectors that span the transverse space.
Let $N_k$, with $1 \leq k \leq d+1-N$, be an orthonormal basis of this
orthogonal space. The $\SO(d,2)$ Lorentz transformations in
Eq.\ \eqref{eq:deltaZ}, with $Z_i = \tilde I$ and $Z_j = N_k$ for any
$1\leq k \leq d+1-N$, leaves all $X_j \, (1\leq j \leq N)$ invariant,
since it only acts in the transverse space. So we obtain a valid
unitarity-compatible IBP-generating vector
\begin{equation}
u(\tilde I, N_k) = \sum_A \biggl[ (\tilde I\, Y_A) \, \left(N_k \, \partialdisplay {Y_A} \right)
     -  (N_k\, Y_A) \, \left(\tilde I \, \partialdisplay {Y_A} \right) \biggr] \,, 
\label{eq:UtildeINk}
\end{equation}
following the notation of the one-loop version in
Eq.~\eqref{eq:uxixj}. However, the IBP relation from the multi-loop
version of Eq.\ \eqref{eq:embedLI},
\begin{equation}
0= \int \biggl( \prod_A \frac{d^{d+2} Y_A \, \delta (Y_A^2 /2 )}{\operatorname{Vol}(\GL(1))} \biggr) 
u(\tilde I, N_k) \,  \mathcal I \,, 
\label{eq:embedLI1}
\end{equation}
breaks the Lorentz symmetry in the $d-(N+1)$ dimensional transverse
space, since it introduces vectors $N_k$ not present in the original
problem, so it is not ideal.  A remedy is to contract the
Lorentz indices to give IBP-generating vectors that are invariant under the
Lorentz symmetry of the transverse directions. We can write down the
following $L$ different vectors,
\begin{equation}
\frac{1}{(-Y_B I)} (N_k Y_B) u(\tilde I, N_k) = 
u \biggl(\tilde I, \frac{\tilde Y_{B\perp}}{(-Y_B I)} \biggr) \,, \label{eq:uIYBperp}
\end{equation}
where the index $k$ is summed and $1 \leq B \leq L$ specifies one of the
independent loop momentum. The label $B$ is not summed
in \eqn{eq:uIYBperp}.  The contraction over the index $k$ ensures
Lorentz invariance in the transverse directions, while the
normalization factor $1/(-Y_B I)$ ensures $\GL(1)$ gauge invariance.
$Y_{B\perp}$ is the projection of $Y_B$ onto the transverse space,
using the inverse of the $(N+1) \times (N+1)$ Gram matrix $g_{ij} =
(X_i X_j)$ with $X_{N+1}\equiv I$,
\begin{equation}
Y_{B\perp}^a = (N_k Y_B) N_k^a = Y_B^a - (Y_B X_i) g^{-1}_{ij} X_j^a \, . \label{eq:defOfBPerp}
\end{equation}
This results in the IBP relations (see \eqn{eq:ibpAsDivergence} for the one-loop analog),
\begin{equation}
0= \int \biggl( \prod_A \frac{ d^{d+2} Y_A \, \delta (Y_A^2 /2 )}{\operatorname{Vol}(\GL(1))} \biggr)
\sum_A \biggl[ (\tilde I\, Y_A) \, \left(N_k \, \partialdisplay {Y_A} \right) -  (N_k\, Y_A) \, 
  \biggl(\tilde I \, \partialdisplay {Y_A} \biggr) \biggr] 
 \biggl( \frac{(N_k Y_B)}{(-Y_B I)}\mathcal I \biggr) , 
\label{eq:embedLI2}
\end{equation}
where there is implicit summation over $k$, and $B$ is a fixed loop label $1,2,\dots,L$.

The right hand side of Eq.\ \eqref{eq:uIYBperp} is an example of an IBP-generating
vector defined using reference vectors with dependence on loop momenta.
The IBP relation from such a vector is a superposition of
familiar $\SO(d,2)$ Lorentz symmetry identities, as in
Eq.\ \eqref{eq:embedLI2}. IBP relations are  obtained
from the vector in explicit components,
\begin{equation}
u\biggl( \tilde I, \frac{\tilde Y_{B\perp}}{(-Y_B I)} \biggr) 
= \biggl(\frac{(\tilde I Y_A)}{(-Y_B I)} \tilde Y_{B\perp}^a 
      - \frac{(\tilde Y_{B\perp} Y_A)}{(-Y_B I)} \tilde I^a \biggr) 
        \partialdisplay{Y_A^a} \,,
\end{equation}
then calculating the total divergence, as in
Eq.\ \eqref{eq:ibpAsDivergence}.  As before, in this expression $A$ is
summed over but $B$ is not.

Since the IBP relations we derived earlier already suffice to reduce
the triangle integrals to bubble integrals, we do not need the
additional IBP relations coming from the transverse space.\footnote{
These additional IBP relations in fact vanish
on the maximal cut, for the three different triangle integrals
considered in the previous section.} But these relations are
needed at the two-loop level.

\subsection{Global and loop-by-loop conformal transformations}
\label{subsec:generalities}

Now consider Lorentz transformations in the embedding space that
affect the external momenta.  To simplify the discussion we focus on
two loops.  We trivially extend the definition of the infinitesimal
Lorentz transformation in Eq.\ \eqref{eq:uxixj} to simultaneously transform both $Y_1$
and $Y_2$,
\begin{equation}
u_{12}(Z_i, Z_j) = \sum_{A=1}^2 u_A(Z_i, Z_j)= 
\sum_{A=1}^2 (Z_{[i}\, Y_A) \, \biggl(Z_{j]} \, \partialdisplay {Y_A} \biggr) \,. 
\label{eq:uxixjTwo}
\end{equation}
Similarly, we will define loop-by-loop Lorentz transformations, namely
\begin{equation}
u_{1}(Z_i, Z_j), \quad u_{2}(Z_i, Z_j)
\end{equation} acting only on $Y_1$ and only on $Y_2$, respectively. For
appropriate $Z_i$ and $Z_j$, $u_{12}(Z_i, Z_j)$ can be considered a
global $\SO(d,2)$ transformation (instead of acting only on the loop
momentum points) that leaves all the external momenta invariant, so that
Eq.~\eqref{eq:uxixjTwo} is a two-loop IBP-generating vector that
does not lead to propagators raised to higher powers. The situation is
entirely analogous to the one-loop case, and allows one-loop
IBP-generating vectors to be reused at higher loops. A difference from
the one-loop case is that we need the IBP-generating vectors arising
from transverse directions, as explained in Subsection
\ref{subsec:boostTrans}, which may be considered as
loop-momentum-dependent global conformal transformations.

For some of the more complicated two-loop integral topologies such as
the penta-box discussed in Subsection \ref{subsec:pentabox},
IBP-generating vectors from global conformal transformations are
\emph{not} sufficient. To deal with this, we construct a class 
of loop-by-loop unitarity-compatible IBP-generating vectors. Consider the
inverse propagators,
\begin{equation}
-\frac{(Y_1 X_i)}{(Y_1 I)}, \quad -\frac{(Y_2 X_j)}{(Y_2 I)}, \quad 
\frac{(Y_1 Y_2)}{(Y_1 I)(Y_2 I)}, \quad \text{with } i \in \sigma_1, \quad j \in \sigma_2\,,
\end{equation}
where $\sigma_1$ and $\sigma_2$ are both subsets of $\{1,2,\dots,N\}$.
If an $\SO(d,2)$ transformation parametrized by the antisymmetric
matrix $\omega_{(1)}^{ij}$ leaves all the $X_i$ points ($i\in
\sigma_1$) invariant, the action of the transformation on $Y_1$ alone
gives the IBP-generating vector
\begin{equation}
V_1^a \partialdisplay{Y_1^a} = \frac 1 2 \omega_{(1)}^{ij} u_1 (X_i, X_j)\,, 
\label{eq:leftVec}
\end{equation}
which does not raise the power of any propagator denominator of the
form $-(Y_1 X_i) / (Y_1 I)$. The vector also does not raise
the power of any propagator denominator of the form $-(Y_2 X_j) / (Y_2
I)$ because the vector does not involve derivatives w.r.t.\ the second
loop momentum. However, the vector may double the power of the
propagator denominator $\D_c \equiv -(Y_1 Y_2) / (Y_1 I) (Y_2 I)$, so
this is not yet a valid unitarity-compatible vector.

Similarly, if a conformal transformation parametrized by $\omega_{(2)}^{ij}$ leaves all the $X_j$ points with $j \in \sigma_2$ invariant, we can write down an IBP-generating vector
\begin{equation}
V_2^a \partialdisplay{Y_2^a} = \frac 1 2 \omega_{(2)}^{ij} u_2 (X_i, X_j)\,, 
\label{eq:rightVec}
\end{equation}
which again does not increase the power of any propagator denominator except for $\D_c$. 
Our  final IBP-generating vector, to be denoted by $\operatorname{cross}(V_1, V_2)$, is
\begin{align}
\operatorname{cross}(V_1^a\partial_{1\, a},V_2^a\partial_{2\, a}) &= \frac{1}{(Y_1 I)(Y_2 I)} \left\{ \left[ V_1^b \partialdisplay{Y_1^b} (Y_1 Y_2) \right] V_2^a \partialdisplay{Y_2^a} - \left[ V_2^b \partialdisplay{Y_2^b} (Y_1 Y_2) \right] V_1^a \partialdisplay{Y_1^a} \right \} \nn \\
&=\frac{1}{(Y_1 I)(Y_2 I)} \left\{ (V_1 Y_2) V_2^a \partialdisplay{Y_2^a} - (V_2 Y_1) V_1^a \partialdisplay{Y_1^a} \right\}, \label{eq:crossVec}
\end{align}
where the overall prefactor
$1/((Y_1 I)(Y_2 I))$ is needed for $\GL(1)$ gauge invariance.
This is designed to annihilate $(Y_1 Y_2)$.  As a
result, this IBP-generating vector does not raise the power of the
propagator denominator $\D_c$. To see this, in Eq.\ \eqref{eq:tangent}
we have
\begin{equation}
\mathcal W_c = \frac{\operatorname{cross}(V_1^a\partial_{1\, a},
         V_2^a\partial_{2\, a}) \rho_c}{\rho_c} 
= - \frac{\operatorname{cross}(V_1^a\partial_{1\, a},
         V_2^a\partial_{2\, a}) \big( (Y_1 I)(Y_2 I) \big)}{(Y_1 I)(Y_2 I)} \,,
\end{equation}
which evaluates to an expression with polynomial dependence on loop
momenta, because the gauge $(Y_1 I) = (Y_2 I) =1$ eliminates the
denominators.

\subsection{The triangle-box}
\label{subsec:tbox}

\begin{figure}
  \centering
  \includegraphics[width=0.4\textwidth]{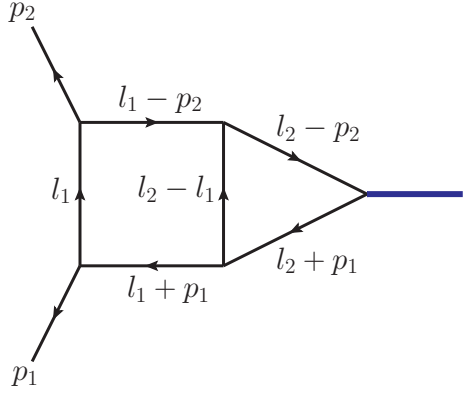}
  \vskip -.3 cm 
  \caption{The triangle-box diagram.}
  \label{fig:tbox}
\end{figure}

As an explicit example, consider the two-loop triangle-box diagram shown in Fig.~\ref{fig:tbox}.
The inverse propagators are
\begin{align}
\D_1 &= (l_1+p_1)^2, \hskip 1 cm \D_2 = l_1^2, \hskip 2.3 cm \D_3 = (l_1 - p_2)^2, \nn \\
\D_4 &= (l_2+p_1)^2, \hskip 1 cm \D_5 = (l_2 - p_2)^2, \hskip 1 cm \D_6 = (l_2-l_1)^2 \,,
\end{align}
while an ``irreducible numerator'', which cannot be written as a linear combination of inverse propagators, is
\begin{equation}
\D_7 = l_2^2 \, .
\end{equation}
Notice that $l_2$ is not the momentum of any propagator, due to our
choice of momentum routing.  The external kinematic invariants are
identical to those of the one-loop triangle with one external mass in
\sect{sec:triangle},
\begin{equation}
p_1^2 = p_2^2 = 0, \quad (p_1+p_2)^2 = s \, .
\end{equation}
Introducing dual coordinates as usual, the six inverse propagators and one irreducible numerator becomes
\begin{align}
\D_1 &= (y_1-x_1)^2, \quad \D_2 = (y_1-x_2)^2, \quad \D_3 = (y_1 - x_2)^2, \nn \\
\D_4 &= (y_2-x_1)^2, \quad \D_5 = (y_2 - x_3)^2, \quad \D_6 = (y_2-y_1)^2, \quad \D_7 =(y_2-x_2)^2 \,,
\end{align}
with the kinematic invariants written as
\begin{equation}
(x_2 - x_1)^2 = (x_3 - x_2)^2 = 0, \quad (x_3 - x_1)^2 = -s \, .
\end{equation}

The triangle-box integral, with the irreducible numerator $\D_7$ raised to the $m$-th power, is written as
\begin{align}
I_m^{\rm tri\operatorname{-}box} &= \int d^d l_1 \int d^d l_2 \frac{\D_7^m}{\D_1 \D_2 \D_3 \D_4 \D_5 \D_6} \nn \\
&= \int \frac{d^{d+2} Y_1 \, \delta (Y_1^2 /2 )}{\operatorname{Vol}(\GL(1))}
\int \frac{d^{d+2} Y_2 \, \delta (Y_2^2 /2 )}{\operatorname{Vol}(\GL(1))} \nn \\
&\quad \times  \frac{(-1)^{6+m} (Y_2 X_2)^m}{(Y_1 I)^{d-4} (Y_2 I)^{d-3+m} (Y_1 X_1) (Y_1 X_2) (Y_1 X_3) (Y_2 X_1) (Y_2 X_3) (Y_1 Y_2)} \, .
 \label{eq:triBoxDef}
\end{align}
Since the external momenta are identical to those for the one-loop triangle, the same subgroup of conformal transformations in dual space leaves the external momenta invariant. Therefore, we can reuse the IBP-generating vectors for the one-loop triangle. The IBP generating vector is parametrized as
\begin{equation}
\frac 1 2 \omega^{ij} u_{12}(X_i, X_j) = \sum_{A=1}^2 \omega^{ij} (X_i Y_A) X_j^a \partialdisplay{Y_A^a} \,,
 \label{eq:omegaVec2}
\end{equation}
which differs from the corresponding one-loop expression Eq.\ \eqref{eq:omegaVec} only by an additional summation over the loop label $A$. We reuse the first solution $\omega_{(1)}$ for the antisymmetric matrix $\omega^{ij}$ in Eq.\ \eqref{eq:triangleOmega} found at one loop. The action of the IBP-generating vector is a straightforward generalization of the one-loop expression Eq.\ \eqref{eq:omegaVecOnYX},
\begin{align}
\Delta_{\omega(1)} (Y_A X_1) &= \Delta_{\omega(1)} (Y_A X_3) = 0, \nn \\
\Delta_{\omega(1)} (Y_A X_2) &= s (Y_A X_2), \nn \\
\Delta_{\omega(1)} (Y_A I) &= -\left[ (Y_A X_1) - 2 (Y_A X_2) + (Y_A X_3) \right] - s (Y_A I), \nn \\
\Delta_{\omega(1)} (Y_1 Y_2) &= 0 \, .
\end{align}
Other than the appearance of the loop label $A$ which may be either $1$ or $2$, the only difference from the one-loop expression is the last line, namely the trivial statement that $(Y_1 Y_2)$ is invariant under simultaneous Lorentz transformations of $Y_1$ and $Y_2$.

IBP relations can be computed in a way similar to how it is done at one loop in Eq.\ \eqref{eq:ibpTriAgain}, in terms of the non-vanishing components of the first solution for $\bar \omega$ in Eq.\ \eqref{eq:triangleOmegaBar},
\begin{align}
0 &= \int \frac{d^{d+2} Y_1 \, \delta (Y_1^2 /2 )}{\operatorname{Vol}(\GL(1))} \int \frac{d^{d+2} Y_2 \, \delta (Y_2^2 /2 )}{\operatorname{Vol}(\GL(1))} \nn \\
&\quad \times  \Delta_{\omega(1)} \frac{(-1)^{6+m} (Y_2 X_2)^m}{(Y_1 I)^{d-4} (Y_2 I)^{d-3+m} (Y_1 X_1) (Y_1 X_2) (Y_1 X_3) (Y_2 X_1) (Y_2 X_3) (Y_1 Y_2)} \nn \\
&= \int\frac{d^{d+2} Y_1 \, \delta (Y_1^2 /2 )}{\operatorname{Vol}(\GL(1))} \int \frac{d^{d+2} Y_2 \, \delta (Y_2^2 /2 )}{\operatorname{Vol}(\GL(1))} \nn \\
&\quad \times \frac{(Y_2 X_2)^m }{(Y_1 I)^{d-4} (Y_2 I)^{d-3+m} (Y_1 X_1) (Y_1 X_2) (Y_1 X_3) (Y_2 X_1) (Y_2 X_3) (Y_1 Y_2)} \nn \\
&\quad \times \bigg\{ (-1)^{6+m} \left[ -\bar \omega_2^{\ 2} (m-1) +(d-4) \bar \omega_4^{\ 4} +(d-3+m) \bar \omega_4^{\ 4} \right] \nn \\
&\quad \left. + (-1)^{5+m} \left[ -\frac{(d-4)}{(Y_1 I)} \left( \sum_{i=1}^3 \bar \omega_4^{\ i} (Y_1 X_i) \right) -\frac{(d-3+m)}{(Y_2 I)} \left( \sum_{i=1}^3 \bar \omega_4^{\ i} (Y_2 X_i) \right) \right] \right\} \, . \label{eq:ibpTBox}
\end{align}
It is illuminating to look at Eq.\ \eqref{eq:ibpTBox} on the maximal cut of the triangle-box, which sets
\begin{equation}
(Y_1 X_1)= (Y_1 X_2)= (Y_1 X_3)= (Y_2 X_1)= (Y_2 X_3)= (Y_1 Y_2)=0 \, . \label{eq:tboxMaxCut}
\end{equation}
After translating Eq.\ \eqref{eq:ibpTBox} back to $\SO(d-1,1)$ loop-momentum space, imposing the maximal cut, and substituting
$\bar \omega_i^{\ j}$ for their explicit values, we obtain
\begin{equation}
0= 2(d-4+m)s\, I_m^{\rm tri\hbox{-}box} + 2(d-3+m) I_{m+1}^{\rm tri\hbox{-}box} + \text{ daughter integrals}, \label{eq:tboxRecurse}
\end{equation}
using the notation of Eq.\ \eqref{eq:triBoxDef} and ``daughter
integrals'' refer to integrals where some of the triangle-box
propagators are canceled. This is a recursion relation which reduces
all the triangle-box integrals to the scalar integral $I_0^{\rm
  tri\operatorname{-}box}$ and integrals with canceled propagators.

We will further show that the scalar triangle-box integral is also reducible to integrals with canceled propagators, by constructing another IBP relation using transformations in the transverse directions as explained in Subsection \ref{subsec:boostTrans}. We define
\begin{equation}
\tilde I = X_2\,,
\end{equation}
which satisfies Eq.\ \eqref{eq:tildeIconditions}
and also define $Y_{1\perp}$ according to Eq.\ \eqref{eq:defOfBPerp} with $B$ set to $1$,
\begin{equation}
Y_{1\perp}^a = Y_1^a - (Y_1 X_i) g^{-1}_{ij} X_j^a \,,
\end{equation}
which is the projection of $Y_1$ onto the transverse space orthogonal to $X_1, X_2, X_3, I$. Using the IBP-generating vector Eq.\ \eqref{eq:uIYBperp} with $B=2$,
\begin{equation}
\frac{-1}{(Y_1 I)} u_{12}(\tilde I, \tilde Y_{1 \perp})\,,
\end{equation}
the IBP relations can be written down as a total divergence as in Eq.\ \eqref{eq:ibpAsDivergence} (but generalized to more than one loops by trivially adding a summation over loop labels $1$ and $2$), with $\mathcal I$ set to
\begin{equation}
\mathcal I = \frac{ (-1)^{6} s}{(Y_1 I)^{d-4} (Y_2 I)^{d-3} (Y_1 X_1) (Y_1 X_2) (Y_1 X_3) (Y_2 X_1) (Y_2 X_3) (Y_1 Y_2)} \, .
\end{equation}
Explicit calculation gives the IBP relation, again dropping terms that vanish on the maximal cut for the purpose of illustration,
\begin{equation}
0 = - (d-3) s \, I_0^{\rm tri\operatorname{-}box} + \hbox{daughter integrals}\, .
\end{equation}
Combined with the recursion relation Eq.\ \eqref{eq:tboxRecurse}, this shows that all triangle-box integrals can be reduced to zero on the maximal cut. In other words, all these integrals can be reduced to integrals with canceled propagators, if we 
retain all terms proportional to inverse propagators in the calculation of the IBP relations. 

\subsection{The double box}
\label{subsec:dbox}

Consider now the two-loop double-box integral in \fig{fig:dbox-dual}.
The inverse propagators with the assigned momentum labels are
\begin{align}
\D_1 &= l_1^2\,, \hskip 1.2 cm 
\D_2 = (l_1 - p_1)^2\,, \hskip 1.2 cm 
\D_3 = (l_1 - p_1 - p_2)^2\,, \hskip 1.2  cm
\D_4 = (l_2 - p_1 - p_2)^2\,, \nonumber \\
\D_5 & =  (l_2 + p_4)^2, \hskip 1.2 cm
\D_6 = l_2^2\,, \hskip 1.2 cm
\D_7 = (l_2 - l_1)^2 \,,
\end{align}
while a choice of irreducible numerators is
\begin{align}
\D_8 &= (l_1 + p_4)^2\,,   \hskip 1.5 cm
\D_9 = (l_2 - p_1)^2 \, .
\end{align}

To write every inverse propagator in the dual-space form, as either $(y_1-y_2)^2$ or $(y_A - x_j)^2$,
we define the following $\SO(d-1,1)$ dual coordinates $x_j$ and $y_A$ such that
\begin{align}
x_2^\mu - x_1^\mu &= p_1^\mu\,, \hskip 1 cm 
x_3^\mu-x_2^\mu = p_2^\mu\,, \hskip 1 cm 
x_4^\mu -x_3^\mu = p_3^\mu\,, \hskip 1 cm
x_1^\mu -x_4^\mu = p_4^\mu\,, \nn\\
y_1^\mu - x_1^\mu &= l_1^\mu\,, \hskip 1 cm
y_2^\mu - x_1^\mu = l_2^\mu \, .
\end{align}
under which the seven inverse propagators become
\begin{align}
\D_1 &= (y_1-x_1)^2\,, \hskip 1 cm
\D_2 = (y_1-x_2)^2\,, \hskip 1 cm
\D_3 = (y_1-x_3)^2\,, \hskip 1 cm
\D_4 = (y_2-x_3)^2\,, \nn \\
\D_5 &= (y_2-x_4)^2\,, \hskip 1 cm
\D_6 = (y_2-x_1)^2\,, \hskip 1 cm
\D_7 = (y_2 - y_1)^2\,,
\end{align}
and the two irreducible numerators become
\begin{equation}
\D_8 = (y_1 - x_4)^2\,, \hskip 1 cm \D_9 = (y_2 - x_2)^2 \,.
\end{equation}
A convenient visualization of the dual points is shown in
Fig.~\ref{fig:dbox-dual}. In terms of these quantities, we define the planar double-box integrand as
\begin{equation}
\Omega_1^{\P} = d^dl_1 d^dl_2 \frac{s t}{\D_1...\D_7}\,. \label{eq:PlanarIntegrand}\\
\end{equation}
where
\begin{equation}
s = (p_1 + p_2)^2 = (x_1 - x_3)^2\,, \hskip 2 cm 
t = (p_2 + p_3)^2 = (x_2 - x_4)^2\,,
\end{equation}
are Mandelstam invariants needed to cancel overall conformal weights.

As usual for planar integrals, we map the dual coordinates $y_A^\mu$
and $x_j^\mu$ to $\SO(d,2)$ embedding-space points $Y_A$ and $X_j$,
following Eqs.\ \eqref{eq:lmap} and \eqref{eq:qmap}. If we use the
algorithm presented in \sect{subsec:embedTri} to find infinitesimal
$\SO(d,2)$ Lorentz transformations that leave all $X_j$ invariant, we
find two such transformations in the notation of
Eq.\ \eqref{eq:uxixj}:
\begin{equation}
u_{12}(X_1, X_3), \quad \quad  u_{12}(X_2,X_4) \,,
\label{eq:dboxIBP1324}
\end{equation}
following the notation of \eqn{eq:uxixjTwo}.
For the one-loop box diagram, both transformations vanish on the
maximal cut because $(YX_i)=0, \, 1\leq i \leq 4$. But for the
two-loop double box topology, $(Y_1 X_4)$ and $(Y_2 X_2)$ are
proportional to the irreducible numerators, so the IBP-generating
vector,
\begin{equation}
u_{12}(X_2, X_4)\,,
 \label{eq:dboxIBPa}
\end{equation}
still gives an IBP-generating vector that does not vanish on the
maximal cut. Eq.~\eqref{eq:dboxIBPa} is essentially the same as the
first IBP-generating vector for the double box in
Ref.~\cite{KosowerNoDouble} obtained using computational algebraic
geometry.  There is another IBP-generating vector for the double box
following the discussion of Subsection \ref{subsec:boostTrans}. We
define another $\SO(d,2)$ embedding-space point $Y_{1\perp}$,
\begin{equation}
Y_{1\perp}^a = Y_1^a - (Y_1 X_i) g^{-1}_{ij} X_j^a\,,  \label{eq:Y1perp}
\end{equation}
where as usual, $g_{ij} = (X_i X_j)$ is the embedding-space Gram matrix, with $X_5$ identified with $I$.
We also define the $\SO(d,2)$ embedding-space point $\tilde I$,
\begin{equation}
\tilde I^a = s (X_2^a + X_4^a) + t (X_1^a + X_3^a) + st \, I^a \,, \label{eq:X0dbox}
\end{equation}
which is the same as Eq.\ \eqref{eq:projectIperpXi} but with an extra overall factor $st$, and satisfies Eq.\ \eqref{eq:tildeIconditions}. Using the IBP-generating vector Eq.\ \eqref{eq:uIYBperp} with $B=1$
we have,
\begin{equation}
u_{12}(\tilde I, Y_{1\perp}) \, . \label{eq:dboxIBPb}
\end{equation}

We have checked using computer algebra that the two IBP-generating vectors, Eqs.\ \eqref{eq:dboxIBPa} and \eqref{eq:dboxIBPb}, with all possible choices of numerators in $\mathcal I$ in the two-loop generalization of Eq.\ \eqref{eq:ibpAsDivergence}, generate a complete set of IBP relations that reduce all double box tensor integrals to two double box master integrals and daughter integrals (i.e.\ integrals with canceled propagators). It is worth noting that the two vectors we found are written down in a very compact form, whereas in Ref.~\cite{KosowerNoDouble} nearly one page is needed to display the vectors found from computational algebraic geometry.

\subsection{The penta-box}
\label{subsec:pentabox}

\begin{figure}
  \centering
  \includegraphics[width=0.43\textwidth]{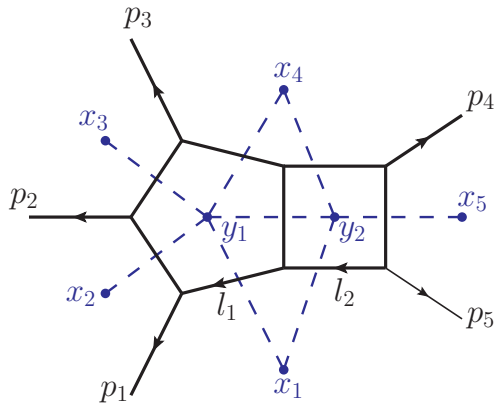}
  \caption{The penta-box integral.}
  \label{fig:penta-box}
\end{figure}

As a more complex example consider the two-loop five-point penta-box
shown in \fig{fig:penta-box}, along with dual coordinates $x_i$ and $y_i$ introduced as usual. 
There are five independent kinematic variables, which may be chosen as
\begin{equation}
s_{12}, s_{23}, s_{34}, s_{45}, s_{51} \,,
\end{equation}
where $s_{ij} = (p_i + p_j)^2$.  The embedding-space Gram matrix is,
identifying $I$ with $X_6$,
\begin{equation}
g_{ij} = (X_i X_j) =
 \begin{pmatrix}
 0 & 0 & -s_{12} & -s_{45} & 0 & 1 \\
 0 & 0 & 0 & -s_{23} & -s_{51} & 1 \\
 -s_{12} & 0 & 0 & 0 & -s_{34} & 1 \\
 -s_{45} & -s_{23} & 0 & 0 & 0 & 1 \\
 0 & -s_{51} & -s_{34} & 0 & 0 & 1 \\
 1 & 1 & 1 & 1 & 1 & 0
 \end{pmatrix} .
\end{equation}
With $\tilde I$ defined as in Eq.\ \eqref{eq:projectIperpXi}, we obtain two IBP-generating vectors from conformal transformations in transverse directions, by setting $B=1,2$ in Eq.\ \eqref{eq:uIYBperp},
\begin{align}
& u \biggl(\tilde I, \frac{\tilde Y_{1\perp}}{(-Y_1 I)} \biggr), \label{eq:pentaboxVec1} \\
& u \biggl(\tilde I, \frac{\tilde Y_{2\perp}}{(-Y_2 I)} \biggr) \, \label{eq:pentaboxVec2} \, .
\end{align}
Next, we examine conformal transformations which do not explicitly involve transverse directions. As in Eq.\ \eqref{eq:ibpAnsatz}, we write down a conformal transformation parametrized as
\begin{equation}
\Delta_\omega Z^a = \sum_{1\leq i,j \leq 6} (Z X_i) \omega^{ij} X_j^a \, ,
\end{equation}
where $\omega$ is a $6\times 6$ antisymmetric matrix. Unlike the previous three-point and four-point examples in this paper, we are not able to find a solution for $\omega^{ij}$ which leaves all external momenta invariant. However, all is not lost. As discussed in the latter half of Subsection \ref{subsec:generalities}, we can look for two different conformal transformations for the two sub-loops, and combine the two to give a unitarity-compatible IBP-generating vector.

We find one solution $\omega^{ij}_{(1)}$ which leaves $x_1, x_2, x_3,
x_4$, or equivalently $p_1,p_2,p_3$, invariant, and three
solutions $\omega^{ij}_{(2a)}, \omega^{ij}_{(2b)}, \omega^{ij}_{(2c)}$
which leaves $x_1,x_4,x_5$, or equivalently $p_4,p_5$ invariant. These
solutions are tabulated in Appendix \ref{sec:pentaboxDetails}.\footnote{In
quoting the number of solutions, we have ignored the solutions which
ultimately do not lead to independent new IBP relations.} Therefore,
the following IBP-generating vectors do not increase the power of any
propagator except the vertical central propagator in
Fig.\ \ref{fig:penta-box},
\begin{align}
&\frac 1 2 \omega^{ij}_{(1)} u_1(X_i,X_j), \quad \frac 1 2 \omega^{ij}_{(2a)} u_2(X_i,X_j), \nn \\
&\frac 1 2 \omega^{ij}_{(2b)} u_2(X_i,X_j), \quad \frac 1 2 \omega^{ij}_{(2c)} u_2(X_i,X_j) \, . \label{eq:pentaBoxSubLoopVecs}
\end{align}
These vectors can be combined to give IBP-generating vectors that do
not increase the power of any propagator. Using the notation of
Eqs.\ \eqref{eq:leftVec}, \eqref{eq:rightVec} and \eqref{eq:crossVec}, these IBP-generating vectors are
\begin{align}
\operatorname{cross} \left( \frac 1 2 \omega^{ij}_{(1)} u_1(X_i,X_j), \frac 1 2 \omega^{ij}_{(2a)} u_2(X_i,X_j) \right), \nn \\
\operatorname{cross} \left( \frac 1 2 \omega^{ij}_{(1)} u_1(X_i,X_j), \frac 1 2 \omega^{ij}_{(2b)} u_2(X_i,X_j) \right), \nn \\
\operatorname{cross} \left( \frac 1 2 \omega^{ij}_{(1)} u_1(X_i,X_j), \frac 1 2 \omega^{ij}_{(2c)} u_2(X_i,X_j) \right) . \label{eq:pentaboxVec345}
\end{align}
We have checked, using computer algebra, that the five IBP-generating
vectors in
Eqs.~\eqref{eq:pentaboxVec1}, \eqref{eq:pentaboxVec2}
and \eqref{eq:pentaboxVec345} are sufficient to reduce all penta-box
integrals to three master integrals.  Again, the five vectors are given
by compact analytic expressions, in contrast to lengthy expressions
one generally finds using computational algebraic geometry.

This formalism generalizes straightforwardly, e.g.\ to the six-point
case, although one would need to check that the IBP relations are
complete for each individual diagram topology, which is left to future work.

\section{Differential equations for planar integrals}
\label{sec:DE}

In this section we briefly comment on applications of the ideas
described in previous sections to constructing differential equations
for integrals. An infinitesimal dual conformal transformation
produces differential equations when we remove the restriction to the
sub-algebra that keeps external legs invariant. We present a treatment
in the embedding space, which simplifies the transformations and has
the advantage that there is no need to fix the translation gauge for
the dual coordinates. In the nonplanar case,  covered in
\sect{sec:nonplanar}, where the transformations for some kinematic
invariants become less obvious it will be simpler to use a ``direct''
treatment.

\begin{figure}
  \centering
  \includegraphics[width=0.32\textwidth]{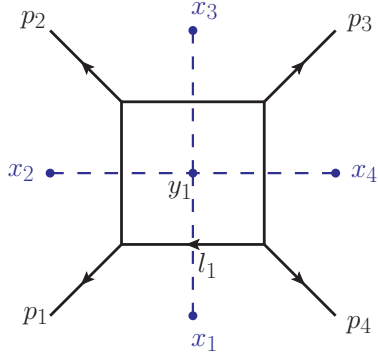}
  \caption{The one-loop box diagram and its dual diagram. 
  }
\vskip -.3 cm 
  \label{fig:box-dual}
\end{figure}

Consider the one-loop box, shown in \fig{fig:box-dual}
which has the same external momenta and the $\SO(d,2)$ points $X_i$ as
the double box in Section \ref{subsec:dbox}. Consider an infinitesimal
$\SO(d,2)$ Lorentz transformation $\Delta Z$ given by
\begin{equation}
\Delta Z^a = (Z_1 Z) Z_2^a - (Z_2 Z) Z_1^a \,,
\end{equation}
with parameters
\begin{equation}
Z_1 = X_2, \quad Z_2 =s \, I + (X_1 + X_3) \,,
\end{equation}
which satisfies
\begin{align}
& (Z_1 X_1) = (Z_1 X_2) = (Z_1 X_3) = 0, \quad (Z_1 X_4) = -t, \quad (Z_1 I) = 1 \nn \\
& (Z_2 X_1)=(Z_2 X_3)=0, \quad (Z_2 X_2)=  (Z_2 X_4) = s, \quad (Z_2 I) = 2 \, .
\end{align}
The transformation of the $\SO(d,2)$ points are
\begin{align}
&\Delta X_1 = \Delta X_3 = 0, \quad \Delta X_2 = -s X_2\,, \nn \\
&\Delta X_4 = -t Z_2 - s Z_1 = -st \, I -t \, X_1 -t\, X_3 -s X_2\,, \nn \\
&\Delta Y=(Y X_2) (s\, I + X_1 + X_3) - s (YI) X_2 - (YX_1) X_2 - (YX_3) X_2\,,
\end{align}
which shows $X_1$, $X_2$, and $X_3$ are invariant up to a $\GL(1)$ gauge scaling. In other words the $d$ dimensional dual coordinates $x_1^\mu$, $x_2^\mu$ and $x_3^\mu$ are left invariant. The factor $(YI)$, which appears in the integration measure, transforms as
\begin{equation}
\Delta (YI) = (I \Delta Y) = 2(YX_2)-(YX_1)-(YX_3)-s(YI)\,.
\end{equation}
As a result, $s=(x_1-x_3)^2$ is invariant, while explicit calculation shows
\begin{equation}
\Delta t = 2(s+t) t \, .
\end{equation}
So the transformation produces differential equations in the $t$ variable,
\begin{align}
2(s+t) t \partialdisplay{t} \left(st I^{\rm box}\right) &= \int \frac{d^{d+2} Y \delta(Y^2/2)}{\operatorname{Vol}(\GL(1))}\, 
 \Delta \biggl(\frac{(X_1 X_3) (X_2 X_4)}{(YI)^{d-4} (Y X_1) (YX_2) (Y X_3) (YX_4)} \biggr)\nn \\
&= \int \frac{d^{d+2} Y \delta(Y^2/2)}{\operatorname{Vol}(\GL(1))} \, 
\frac{(X_1 X_3) (X_2 X_4)}{(Y X_1) (YX_2) (Y X_3) (YX_4)} \, \Delta \biggl(\frac{1}{(YI)^{d-4}}\biggr) \nn \\
&=  \int \frac{d^{d+2} Y \delta(Y^2/2)}{\operatorname{Vol}(\GL(1))}\, 
\frac{(X_1 X_3) (X_2 X_4)}{(Y X_1) (YX_2) (Y X_3) (YX_4)} \frac{1}{(YI)^{d-4+1}} \nn \\
& \hskip 1.5 cm  \times (-d+4) \left[ 2(YX_2)-(YX_1)-(YX_3)-s(YI) \right] \nn \\
&= \epsilon \left[ -2 s \left(st I^{\rm box}\right) + 4 s t\, I^{{\rm tri},t} - 4 s t \, I^{{\rm tri},s} \right],
\end{align}
where the last line consists of the box, the $t$-channel triangle, and the $s$-channel triangle integrals.
After summing $s$ and $t$ channel versions of this equation it immediately reproduces
Eq.~(4.11) of Ref.~\cite{OneLoopPentagon}.


It is noteworthy that the right hand side of the differential equation
so derived is proportional to $\epsilon$~\cite{HennDifferentialEqs}.
It is perhaps not too surprising that this structure naturally arises
in our approach.  If we ignore the effect of the regulator, the
combination $s t I^{\rm box}$ is invariant under dual conformal
transformations in four dimensions.  However, the box integral is
infrared singular so a regulator is required.  Dimensional
regularization breaks the invariance, so instead of finding zero on
the right-hand-side we find terms proportional to $\epsilon$.  Besides
leading to simpler differential equations, integrals with such
symmetries are expected to have interesting properties, including
uniform transcendentality~\cite{UniformTranscendental} and $d \log$
forms~\cite{DLogForms}.  It would be interesting to further explore
these ideas at higher loops, not only for the planar case, but also for
nonplanar integrals in the context of the approach of \sect{sec:nonplanar}.

We end this section with some discussions about the applicability of
this method to more complicated integral topologies. First, let us
look at the number of legs allowed. In this simple example, a
conformal boost changes the dimensionless ratio of Mandelstam
variables, $s/t$, while a scaling transformation rescales both $s$ and
$t$. Together these two transformations allow the \emph{whole phase
space} of external kinematic invariants to be explored. For massless
planar diagrams, this breaks down when there are six or more external
legs, because nontrivial conformally invariant cross ratios
exist~\cite{CrossRatios}, and conformal transformations only allow us
to explore a subspace of the phase space with the same cross
ratios. 

Second, consider the $\epsilon$ factorization properties of the
differential equations for more general integrals. For any integrand
that is dual conformal invariant, our method automatically leads to
differential equations where there is an explicit factor of $\epsilon$
on the right hand side.  For more complicated examples beyond the
one-loop box, it is generally necessary to perform
unitarity-compatible IBP reduction to bring the right hand side into a
linear combination of master integrals.  Assuming that IBP reduction
of the right hand side does not introduce singularities, this gives a
symmetry-based understanding of Henn's $\epsilon$ form of differential
equations. For planar integrals that are not invariant, we still
obtain differential equations without raised propagator powers. This
allows unitarity-compatible IBP reduction to be used to simplify the
differential equations, even though we no longer would have $\epsilon$
factorization prior to IBP reduction. Third, the applicability of our
method to nonplanar topologies will be demonstrated in the next
section, where differential equations are derived for the nonplanar
double box by identifying a symmetry analogous to dual conformal
symmetry.

\section{Nonplanar analog of dual conformal symmetry}
\label{sec:nonplanar}

In this section we find a nonplanar analog of dual conformal
transformations at two loops.  We do so by working out the symmetries
of two-loop integrals with three or four external legs.

\subsection{Hidden symmetry of a two-loop nonplanar three-point integral}

\begin{figure}
  \centering
  \includegraphics[width=0.5\textwidth]{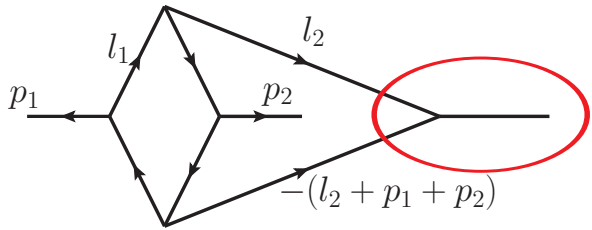}
  \caption{The crossed triangle-box, with two massless legs $p_1$ and
    $p_2$, and one massive leg shown as a thick line.  We remove the
    right-most part of the diagram enclosed in a (red) ellipse, in
    order to open up the diagram into a one-loop planar diagram.}
  \label{fig:np}
\end{figure}

\begin{figure}
  \centering
  \includegraphics[width=0.4\textwidth]{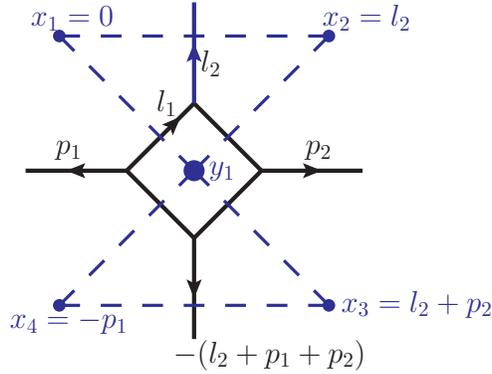}
  \caption{This figure is obtained from Fig.\ \ref{fig:np} by removing
    the rightmost part enclosed in the (red) ellipse, including the
    massive leg. The result is a planar diagram,
    allowing dual coordinates $x_i$ to be introduced. Each of the
    dashed (blue) lines corresponds to one of the six propagators in
    the integral.}
  \label{fig:np-dual}
\end{figure}

We start by deriving IBP-generating vectors for a two-loop
nonplanar integral topology, the crossed triangle-box shown in
\fig{fig:np}, with two massless outgoing external momenta $p_1$
and $p_2$, and one outgoing massive external momentum $-(p_1+p_2)$ on the
right.  The kinematics is given by
\begin{equation}
p_1^2=p_2^2=0, \quad (p_1+p_2)^2 = s \, .
\end{equation}
Our strategy is to open up the nonplanar diagram by removing vertices
in the graph. This strategy has been pursued in
Ref.~\cite{NonplanarYangian} to find symmetries of leading
singularities of nonplanar integrands.  Here we will find symmetries of
the \emph{complete off-shell} integrand, up to anomalies introduced by
dimensional regularization of infrared singularities, similar to the
situation in the planar case. A key hint comes from the fact that in the planar case, dual conformal transformations generate polynomial tangent vectors of unitarity cut surfaces, under which each propagator transforms with a polynomial weight as in Eq.\ \eqref{eq:tangentDE}. Therefore, we will first find transformations of nonplanar integrands with this property, before verifying that such transformations are in fact a symmetry of the integrand with appropriate numerators.

To open up the nonplanar diagram in \fig{fig:np} into a planar one, we
remove the massive external leg and the vertex attached to it enclosed
in the (red) circle, producing a planar one-loop diagram in
Fig.~\ref{fig:np-dual}, in which two ``external'' legs depend on the
second loop momentum.  Dual coordinates may be introduced for this
planar one-loop diagram, as illustrated by the dashed
lines in Fig.~\ref{fig:np-dual}. In this case, we find
it more convenient to directly work with conformal transformations in
$d$ dimensions rather than using the embedding formalism. The inverse
propagators are expressed as squared differences between pairs of
points in dual spacetime,
\begin{align}
& \D_1=l_1^2 = (y_1 - x_1)^2\,, \hskip 3.5 cm   \D_2=(l_1 - l_2)^2 = (y_1 - x_2)^2\,, \nn \\
& \D_3=(l_1-l_2-p_2)^2 = (y_1 - x_3)^2\,, \hskip 1.35 cm 
  \D_4=(l_1 + p_1)^2 = (y_1 - x_4)^2\,, \nn \\
& \D_5=l_2^2 = (x_2 - x_1)^2\,,   \hskip 3.5 cm 
  \D_6=\left[-(l_2+p_1+p_2)\right]^2 = (x_4 - x_3)^2 \,,
\end{align}
while the only irreducible numerator can be chosen as
\begin{equation}
\D_7=(l_2 + p_1)^2 = (x_2 - x_4)^2 \, .
\end{equation}
While the external momentum $p_1$ and $p_2$ each can be written as the difference between two dual coordinates, this is no longer true for the massive external momenta $-(p_1+p_2)$, in contrast to the planar case.
Choosing a gauge $x_1=0$ to fix the translation degree of freedom, the dual coordinates are positioned at
\begin{align}
x_1&= 0, \hskip 1.5 cm  x_2=l_2, \hskip 1.5 cm  y_1 = l_1,  \hskip 1.5 cm 
x_3 = l_2 +p_2, \hskip 1.5 cm  x_4 = -p_1 \, . \label{eq:npDualPointsPos}
\end{align}
Using these variables, the crossed triangle-box integral in
\fig{fig:np}, with $m$ powers of the irreducible numerator, is 
\begin{align}
I_m^{\rm ctb} &= \int d^d y_1 \int d^d x_2 \frac{\D_7^m}{\D_1 \D_2
  \D_3 \D_4 \D_5 \D_6} \nn \\ &= \int d^d y_1 \int d^d x_2 \frac{(x_2
  - x_4)^{2m}}{(y_1 -x_1)^2 (y_1-x_2)^2 (y_1-x_3(x_2))^2 (y_1-x_4)^2
  (x_2 - x_1)^2 (x_4 -x_3(x_2))^2} \,,
\label{eq:Ictb}
\end{align}
where $x_3$ is taken to be a function of $x_2 = l_2$.

The expression in \eqn{eq:Ictb} is in a form where we can
conveniently apply conformal transformations.  An infinitesimal
transformation, consisting of a conformal boost with
parameter $b^\mu$, a scaling with parameter $\beta$, and a Lorentz
transformation $\Omega^\mu_{\ \nu}$, is given by
\begin{equation}
\Delta z^\mu = \frac 1 2 z^2 \, b^\mu + (\beta - b \cdot z) z^\mu + \Omega^\mu_{\ \nu} z^\nu \,.
 \label{eq:conformalTransWithLorentz}
\end{equation}
Under the transformation,  each inverse propagator of the form $(z_1-z_2)^2$ has a weight given by Eq.\ \eqref{eq:DualConfOnInvProp},
\begin{equation}
\left[ 2 \beta - b \cdot (z_1 + z_2) \right] \, .
 \label{eq:propWeight}
\end{equation}
For a propagator given by $1/(z_1-z_2)^2$, the weight has an opposite sign. 
Meanwhile, the integration measures $d^d y_1$ and  $d^d x_2$ have a weight given by Eq.\ \eqref{eq:weightOfDz},
\begin{equation}
\frac{\partial \Delta y_1^\mu}{\partial y_1^\mu} = \left( \beta - b \cdot y_1 \right)d \,,
\hskip 2 cm 
\frac{\partial \Delta x_2^\mu}{\partial x_2^\mu} = \left( \beta - b \cdot x_2 \right)d \, .
\end{equation}
For the nonplanar integral in Eq.\ \eqref{eq:Ictb}, the total weight, from the integration measures, irreducible numerators, and propagators, is
\begin{align}
\mathcal W(m,b,\beta) = & d \left( \beta - b \cdot y_1 \right) +  d \left( \beta - b \cdot x_2 \right) +m \left[ 2 \beta - b \cdot (x_2 + x_4) \right] \nn \\
- & \left( \sum_{i=1}^4 \left[ 2 \beta - b \cdot (y_1 + x_i) \right] \right) -\left[ 2 \beta - b \cdot (x_1+x_2) \right] - \left[ 2 \beta - b \cdot (x_3 + x_4) \right] ,
\end{align}
which, using the explicit expression Eq.\ \eqref{eq:npDualPointsPos}, becomes
\begin{equation}
\mathcal W(m,b,\beta) = 2 \beta(d+m-6) + b \cdot \left[ -(d-4) (l_1 + l_2) + 2p_2 - 2p_1 - m(l_2-p_1) \right] . \label{eq:weightM}
\end{equation}

We obtain IBP-generating vectors when the transformation Eq.\ \eqref{eq:conformalTransWithLorentz} 
leaves both $p_1$ and $p_2$ invariant, i.e.\
\begin{equation}
\Delta p_1 = \Delta x_1 - \Delta x_4 =0\,, \hskip 1.5 cm  \Delta p_2 = \Delta x_3 - \Delta x_2 = 0 \, .
\end{equation}
A solution for such a transformation is
\begin{equation}
b = p_2\,, \hskip 1cm \beta = -\frac{p_1 \cdot p_2} 2 = - \frac s 4\,,
 \hskip 1 cm 
\Omega^\mu_{\ \nu}  = \frac 1 2 p_{1\,\nu}  p_2^\mu - \frac 1 2 p_{2 \, \nu} p_1^\mu \, .
\end{equation}
The weight \eqref{eq:weightM} is then (using $p_2^2=0$),
\begin{align}
\mathcal W(m,p_2,-\frac s 4) &= -\frac s 2 (d+m-6) + p_2 \cdot \left[-(d-4)(l_1+l_2)-2p_1-m\, (l_2 - p_1) \right] \nn \\
&= \frac{1}{2} (d-4+m)s  + \left(d-4+\frac m 2\right) (\D_7 - \D_6) + \frac 1 2 (d-4) (\D_3 - \D_2) \, .
\label{eq:TotalWeight}
\end{align}
Remarkably, the above expression vanishes when $d=4$ and $m=0$. This
shows the integrand of the scalar integral $I^{\rm ctb}_0$ is invariant
under a nontrivial infinitesimal transformation.\footnote{By
  ``nontrivial'', we mean that the transformation is not a Lorentz
  transformation (of both external and loop momenta) which trivially
  leave the integral invariant.}

The IBP relation obtained from \eqn{eq:TotalWeight} is
\begin{align}
0 &= \int d^d l_1 \int d^d l_2 \frac{\D_7^m \mathcal W(m,p_2,-\frac s 4)}{\D_1 \D_2 \D_3 \D_4 \D_5 \D_6} = \frac 1 2 (d-4+m)s \, I_m^{\rm ctb} + \left(d-4+\frac m 2\right) I_{m+1}^{\rm ctb} \nn \\
&\quad + \text{daughter integrals} \, , \label{eq:ctbRecursion}
\end{align}
which reduces all integrals of this topology to the scalar master
integral $I_0^{\rm ctb}$ and daughter integrals with canceled
propagators. We checked that Eq.\ \eqref{eq:ctbRecursion} agrees with
maximal-cut IBP relations obtained from computational algebraic
geometry.  Since this is a single scale integral, differential
equations are not useful unless additional scales are
introduced~\cite{Henn2013nsa}; in any case its value
is given in Ref.~\cite{CrossTri}.

\subsection{Hidden symmetry of two-loop four-point nonplanar integrals}

\begin{figure}
  \centering
  \includegraphics[width=0.5\textwidth]{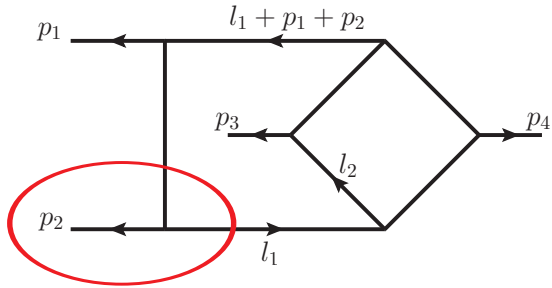}
  \caption{The two-loop nonplanar crossed box. The part of the diagram enclosed
    in a red ellipse will later be removed, so that the diagram is
    broken up into a one-loop planar diagram.}
  \label{fig:npdbox}
\end{figure}

Consider now the two-loop four-point nonplanar integral with massless
external legs displayed in \fig{fig:npdbox}.  In this case, if we
follow the same procedure as for the nonplanar triangle-box, we find
no solution for a generalized dual conformal transformation that
leaves the external points invariant, so the construction does not
generate IBP relations.  However, by relaxing this condition, we have
no difficulty finding an invariance of the integrals with appropriate
numerators.  We use it to construct differential equations along the
lines of \sect{sec:DE}, implying that the symmetry determines the
analytic structure.  As we emphasize in the subsequent section, this
implies that the nonplanar sector of the two-loop four-point $\mathcal
N = 4$ super-Yang--Mills amplitude has a hidden symmetry analogous to
dual conformal symmetry.

\begin{figure}
  \centering
  \includegraphics[width=0.45\textwidth]{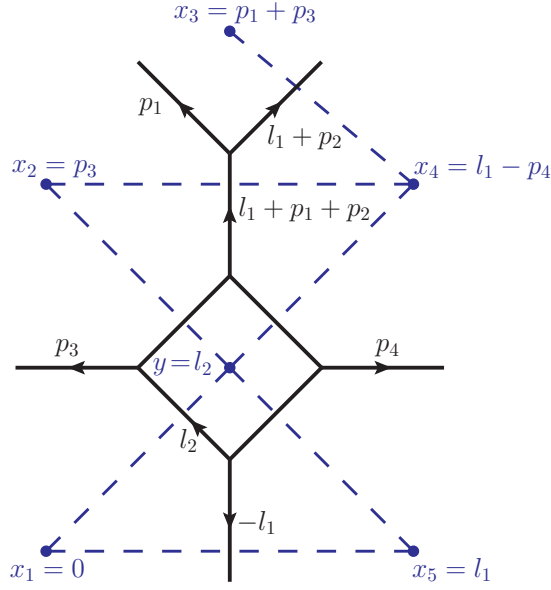}
  \caption{The planar one-loop diagram obtained by removing the bottom left part of Fig.\ \ref{fig:npdbox}. This allows one to introduce dual coordinates $x_i$. Each of the dashed lines corresponds to one of the six propagators in the integral.}
  \label{fig:npdbox-dual}
\end{figure}

In order to define an analog of dual conformal symmetry we open the
diagram by removing the part of the diagram in \fig{fig:npdbox} enclosed by a red ellipse,
including the leg with external momentum $p_2$. This opens up the
two-loop diagram into a one-loop diagram with ``fake'',
loop-momentum-dependent external legs as shown in
Fig.\ \ref{fig:npdbox-dual}.  With this construction every propagator
momentum is expressed as the difference between two dual-space
points. Each of the external momenta $p_1$, $p_3$, and $p_4$ is also
expressed as the difference between two dual coordinates, though the
same is {\it not} true for $p_2$ (in contrast to the planar case).
Although one might worry that this may cause problems with the
construction, we shall see that it does not.

We take the inverse propagators as,
\begin{align}
\D_1&=l_1^2=(x_1-x_5)^2, \quad\quad \D_2 = (l_1+p_2)^2 = (x_4-x_3)^2\,, 
 \quad\quad \D_3 = (l_1+p_1+p_2)^2=(x_4-x_2)^2\,, \nn \\
\D_4 &= (l_1-l_2)^2 = (y-x_5)^2\,, \quad\quad \D_5 = l_2^2 = (y-x_1)^2\,, \nn \\ 
\D_6 & = (l_2-p_3)^2 = (y-x_2)^2\,, \quad\quad  \D_7  = (l_2-l_1+p_4)^2 = (y-x_4)^2 \,,
\end{align}
and we have chosen the gauge 
\begin{equation}
\hskip -.12 cm 
x_1=0\,, \quad\quad x_2 = p_3\, , \quad\quad x_3=p_1+p_3\,, 
\quad\quad x_4= l_1-p_4\,, \quad\quad x_5 =l_1\,, \quad\quad y = l_2 \, .
\end{equation}
We can define two numerators (which are not independent irreducible numerators, nevertheless are convenient for notational purposes),
\begin{equation}
\D_8=(l_1-p_3)^2 = (x_5-x_2)^2, \quad \D_9 = (l_1-p_4)^2 = (x_4-x_1)^2 \, .
\end{equation}
We also
note that
\begin{equation}
u=(p_1+p_3)^2 = (x_3-x_1)^2 \, .
\end{equation}

Refs.~\cite{TwoLoopN4Nonplanar,NonplanarAmplituhedron} expressed the
two-loop four-point amplitude in terms of integrals that
have only logarithmic singularities, reflecting a property of the full
amplitude.  In this representation the two nonplanar integrands that appear
in the amplitude (up to relabelings) are,
\begin{align}
\Omega_1^{\NP} &= d^dl_1 d^dl_2 \frac{s u \D_8}{\D_1...\D_7}\,, \label{eq:symIntegrand1}\\
\Omega_2^{\NP} &= d^dl_1 d^dl_2 \frac{s t \D_9}{\D_1...\D_7} \, . \label{eq:symIntegrand2}
\end{align}
The normalization of each is chosen so it has unit leading
singularity~\cite{NonplanarAmplituhedron}.  
Our task will be to find a hidden symmetry responsible for the 
simple analytic properties after integration.

\begin{figure}
  \centering
  \includegraphics[width=0.4\textwidth]{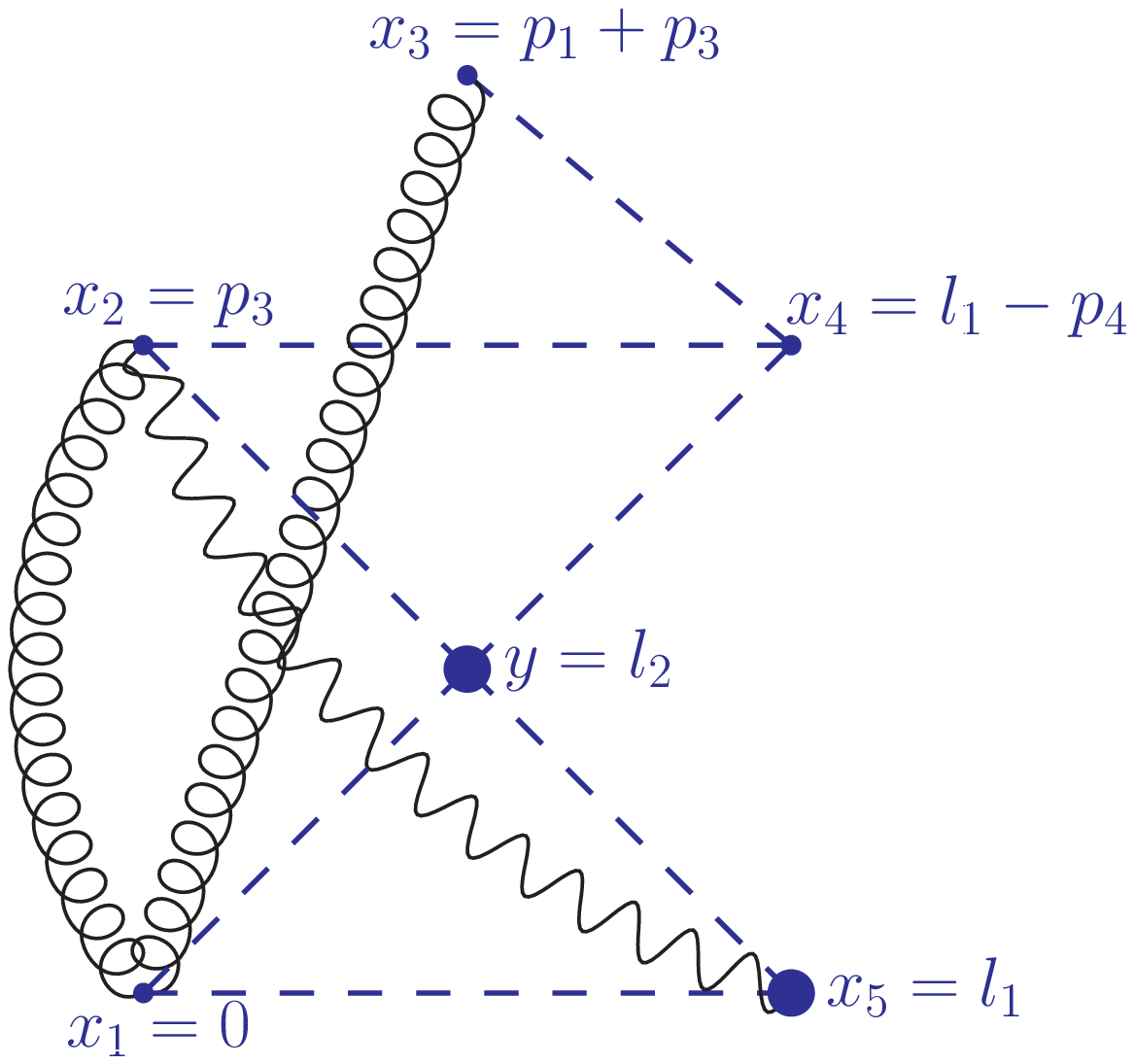}
  \caption{Weight diagram for the integrand \eqref{eq:symIntegrand1} 
    under the conformal boost \eqref{eq:npdbox-trans}.}
  \label{fig:symWeightDiagram1}
\end{figure}

\begin{figure}
  \centering
  \includegraphics[width=0.4\textwidth]{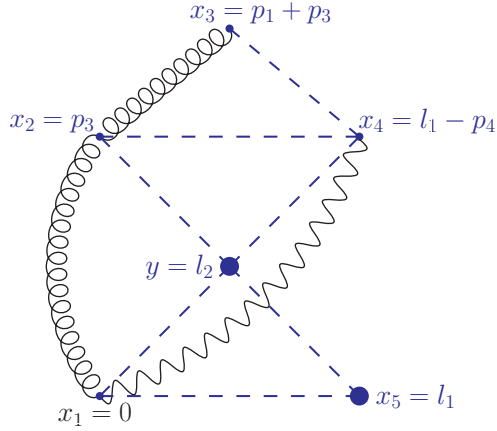}
  \caption{Weight diagram for the integrand \eqref{eq:symIntegrand2} under the conformal boost \eqref{eq:npdbox-trans}.}
  \label{fig:symWeightDiagram2}
\end{figure}

A conformal boost on the $x_i$'s and $y$ with parameter $b^{\mu}$ gives
\begin{align}
\Delta p_4 &\equiv \Delta x_5 -\Delta x_4\\
&=(l_1\cdot p_4)b -(p_4\cdot b) l_1 -(l_1\cdot b-p_4\cdot b) p_4 \, .
\end{align}
The appearance of loop momentum in the transformation of external momentum is not 
surprising, given that once we cut a nonplanar diagram open internal momenta effectively
become ``external''. This is, of course, not desirable if we wish to use the transformations
to construct differential equations. To remove the loop-momentum dependence of this variation, we
simply choose
\begin{equation}
b =p_4 \,.
\label{eq:npdbox-trans}
\end{equation}
While this restricts the transformations to a subset of
conformal transformations, we shall see that this is sufficient for
constructing differential equations analogous to those of the planar
case.  Applying the conformal transformation gives
\begin{align}
\Delta p_1 &= \Delta x_3 - \Delta x_2 = \frac{1}{2}(u\, p_4 - t\, p_3 + u\, p_1)\,,\nn\\
\Delta p_3 &= \Delta x_2 -\Delta x_1 = -\frac{1}{2}s\, p_3 \,,\nn\\
\Delta p_4 &=\Delta x_5 -\Delta x_4 = 0 \,,
\label{eq:NonplanarDoubleBox}
\end{align} 
so that the masslessness of these three external legs is preserved.
In fact, the masslessness of $p_1$, $p_3$, and $p_4$ are trivially preserved by the properties
of conformal transformations, since each of these three momenta is the difference between two points in dual space.
Remarkably, the same is nontrivially true of the second leg, as can be readily checked,
\begin{equation}
\Delta p_2^2 = 2 p_2\cdot \Delta p_2 = -2 p_2\cdot (\Delta p_1 + \Delta p_3 + \Delta p_4) = 0 \,.
\end{equation}
This ensures that the transformation preserves the masslessness of all external legs,
which is essential for the construction to be useful.
In addition we have, 
\begin{align}
\Delta s &= \Delta (2 p_3\cdot p_4) = 2(p_3 \cdot \Delta p_4 +p_4 \cdot \Delta p_3) = -\frac{s}{2}s \,,\nn \\ 
\Delta t &= \Delta (2 p_1\cdot p_4) = 2(p_1 \cdot \Delta p_4 +p_4 \cdot \Delta p_1) = -\frac{t+2s}{2}t\,. \label{eq:npdbox-DeltaST}
\end{align}
Note that applying Eq.\ \eqref{eq:DualConfOnInvProp} directly gives, 
\begin{equation}
\Delta u = -b\cdot (x_3+x_1) u  = -p_4\cdot (p_1+p_3) u 
=-\frac{u}{2}(t+s) =-\Delta s -\Delta t \,,
\label{eq:npdbox-DeltaU}
\end{equation}
which is consistent with momentum conservation.  It will be convenient
for later purposes to write down the weights of $s$, $t$, and $u$
under the transformation, as dot products between $(-p_4)$ and other
momenta,
\begin{align}
\mathcal W_s &\equiv \frac{\Delta s}{s} = -p_4 \cdot p_3 \,, \nn \\
\mathcal W_t &\equiv \frac{\Delta t}{t} = -p_4 \cdot (p_1 + 2p_3)\,, \nn \\
\mathcal W_u &\equiv \frac{\Delta u}{u} = -p_4 \cdot (p_1 + p_3)\,. \label{eq:weightOfSTU}
\end{align}
Meanwhile, a numerator of the form $(z_i-z_j)^2$ has the weight $-p_4
\cdot (z_i + z_j)$, while an extra minus sign is present in the weight
for a propagator of the form $1/(z_i-z_j)^2$. The weight of the
integration measure is given by $-d\, p_4 \cdot (x_5 + y) = -d\, p_4
\cdot (l_1+l_2)$.  We can now straightforwardly prove that the
nonplanar contributions to the $\mathcal N=4$ super-Yang--Mills
amplitudes are invariant under this transformation.  Namely, in $d=4$,
the two integrands Eqs.\ \eqref{eq:symIntegrand1} and
\eqref{eq:symIntegrand2} in the amplitudes transform as,
\begin{align}
\Delta \Omega_1^{\NP} = \Delta \Omega_2^{\NP}=0 \, .
\end{align}

A pictorial way to derive the above equation is as follows. We
have shown that the weights of the Mandelstam variables, numerators,
propagators, and integration measures are each written in the form
$-p_{4} \cdot W$ for some ``weight vector'' $W^\mu$. So it is
convenient to represent the weight of the integrand diagrammatically as
in Fig.~\ref{fig:symWeightDiagram1} and~\ref{fig:symWeightDiagram2}.
In the diagrams, the weight of a propagator of the form
$1/(z_1-z_2)^2$ is represented by a dashed line connecting two points
$z_1$ and $z_2$, contributing $-(z_1^\mu + z_2^\mu)$ to the weight vector
$W^\mu$. The weight of a numerator of the form $(z_1-z_2)^2$ is
represented by a wiggly line connecting two points $z_1$ and $z_2$,
contributing $z_1^\mu + z_2^\mu$ to the weight vector $W^\mu$. The
weight of the Mandelstam variables appearing in the numerator is
represented by a coil-like line connecting two points $z_1$ and $z_2$,
again contributing $z_1^\mu+z_2^\mu$ to the weight vector $W^\mu$. To
reproduce Eq.\ \eqref{eq:weightOfSTU}, for $\mathcal W_s$ we choose
$z_1=0=x_1$ and $z_2=p_3=x_2$, for $\mathcal W_t$ we choose
$z_1=p_3=x_2$ and $z_2=p_1+p_3=x_3$, and for $\mathcal W_u$ we choose
$z_1 = 0 = x_1$ and $z_2 = p_1+p_3 = x_3$. Finally, the weight of the
integration measure is indicated by large black dots at the two points
$x_5$ and $y$. In our notation, a large black dot at any point $z$
contributes $d\, z^\mu$ to the weight vector $W^\mu$, with $d$ being
the spacetime dimension. The total weight vector $\sum W^\mu$ can now
be read off from the diagram in the following manner: at each vertex
(i.e.\ a dual space point) $z^\mu$, we count the number of wiggly
lines and coil-like lines joining the vertex, subtract the number of
dashed lines joining the vertex, and add the spacetime dimension $d$
if a large black dot appears at the vertex. The final number is
multiplied by $z^\mu$ and included in $\sum W^\mu$. For the first
integrand $\Omega_1^{\NP}$ in Eq.\ \eqref{eq:symIntegrand1}, the
weight diagram Fig.\ \ref{fig:symWeightDiagram1} gives the weight
vector
\begin{align}
\sum W_1^\mu &= (2-2)x_1^\mu + (2-2) x_2^\mu + (1-1) x_3^\mu 
- 3 x_4^\mu + (1-2+d) x_5^\mu + (-4+d) y^\mu \nn \\
&= (d-4) (l_1^\mu + l_2^\mu) + 3 p_4^\mu \, .
\end{align}
Using this, we arrive at 
\begin{equation}
\Delta \Omega_1^\NP  = \left( -p_4 \cdot \sum W_1 \right) \Omega_1^\NP = -(d-4) p_4 \cdot (l_1 + l_2) \Omega_1^\NP \, . \label{eq:Omega1finalWeight}
\end{equation}
Since the transformation changes the Mandelstam variables as in Eq.\ \eqref{eq:npdbox-DeltaST}, we arrive at a differential equation for the Feynman integral,
\begin{equation}
\left( -\frac{s^2}{2} \partialdisplay{s} - \frac{t(t+2s)}{2} \partialdisplay{t} \right) \int \Omega_1^\NP 
= -(d-4) \int p_4 \cdot (l_1 + l_2)\, \Omega_1^\NP \, . \label{eq:npdbox-DE1a}
\end{equation}
Since $\Omega_1^\NP$ has mass dimension $2(d-4)$, we trivially obtain another differential equation from the simultaneous scaling of all Mandelstam variables,
\begin{equation}
\frac s 2 \left( s \partialdisplay{s} + t \partialdisplay{t} \right) \int \Omega_1^\NP = \frac s 2 (d-4) \int \Omega_1^\NP  \, . 
\label{eq:npdbox-DE1b}
\end{equation}
Adding Eqs.\ \eqref{eq:npdbox-DE1a} and \eqref{eq:npdbox-DE1b}, we obtain the derivative of the integral against $t$ only
\begin{equation}
\frac{tu}{2} \partialdisplay{t} \int \Omega_1^\NP = \epsilon \left[ 2 \int p_4 \cdot (l_1 + l_2) -s \right] \Omega_1^\NP \, . 
\label{eq:npdbox-DE1}
\end{equation}
For the second integrand $\Omega_2^\NP$ in Eq.\ \eqref{eq:symIntegrand2}, again we read off the weight vector from Fig.\ \ref{fig:symWeightDiagram2}. This leads to results similar to those for $\Omega_1^\NP$,
\begin{equation}
\Delta \Omega_2^\NP  = \left( -p_4 \cdot \sum W_1 \right) \Omega_1^\NP = -(d-4) p_4 \cdot (l_1 + l_2) \Omega_2^\NP \,, 
\label{eq:Omega2finalWeight}
\end{equation}
and
\begin{equation}
\frac{tu}{2} \partialdisplay{t} \int \Omega_2^\NP = \epsilon \left[ 2 \int p_4 \cdot (l_1 + l_2) -s \right] \Omega_2^\NP \,.
\label{eq:npdbox-DE2}
\end{equation}
If there were no infrared singularities, we would be able to set
$\epsilon = 0$ and the symmetry would be exact.

\begin{figure}
  \centering
  \includegraphics[width=0.4\textwidth]{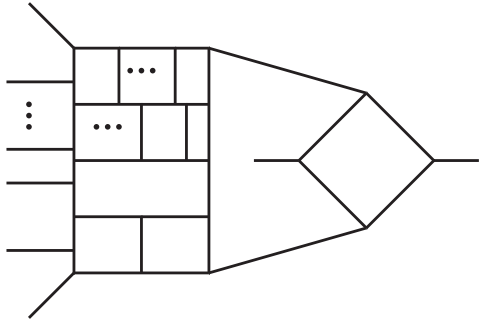}
  \caption{An illustrative multi-loop diagram, where a similar 
   analysis as for the nonplanar double box in \fig{fig:npdbox}
  identifies a hidden symmetry.}
  \label{fig:CrossedMultiloop}
\end{figure}

It is interesting that a similar analysis extends to any nonplanar
diagram with a single crossed box at any loop order, as illustrated in
\fig{fig:CrossedMultiloop}.  In particular, if we consider this
diagram with a numerator obtained from a corresponding planar dual
conformal invariant one, except for the single crossed box which is
given a similar factor as in the two-loop cases
\eqref{eq:symIntegrand1} and \eqref{eq:symIntegrand2}, then the
resulting nonplanar integral possesses a higher-loop analog of dual
conformal symmetry.  A way to show the invariance is to remove a three
vertex from the crossed box and perform a similar analysis to the one
of the previous section for the two-loop crossed box.  In this case,
it is convenient to remove a vertex from the crossed double box,
instead of the other parts of the diagram.  We have checked that the
analog of dual conformal symmetry is present for this class of
nonplanar integrals, at any loop order.  As for the two-loop case, we
can use it to generate differential equations to constrain the
integrals whose right hand side is proportional to the dimensional
regularization parameter,~$\epsilon$.

\section{Invariance of the nonplanar two-loop four-point 
$\mathcal N = 4$ super-Yang--Mills amplitude}
\label{sec:nonplanaramplitude}

In the previous section we identified a new symmetry of the nonplanar
integrands that appear in the two-loop four-point amplitude of
$\mathcal N = 4$ super-Yang--Mills theory.  In this section we
comment on symmetries of the full amplitude.

From Eq. (3.15) of Ref.~\cite{NonplanarAmplituhedron} we have the full 
two-loop four-point amplitude of $\mathcal N = 4$ super-Yang-Mills theory as
\begin{align}
\mathcal{A}_4^{\rm 2\hbox{-}loop} = &  - \frac{g^6}{4 (2\pi)^{2D}} \sum_{{\mathcal S}_4} 
   \Bigr[ c^{\P}_{1234} A^\tree(1,2,3,4) \int \Omega^{\P}
 \label{eq:TwoLoopAmplitude} \\
& \hskip 2.5 cm  \null
       - c^{\NP}_{1234} \Bigl(A^\tree(1,2,4,3) \int \Omega_1^{\NP} +
                A^\tree(1,2,3,4) \int \Omega_2^{\NP} \Bigr) \Bigr]\,, \nn
\end{align}
where $c^{\rm P}_{1234}$ and $c^{\rm NP}_{1234}$ are the planar and
nonplanar color factors obtained by dressing the diagrams in
\figs{fig:dbox-dual}{fig:npdbox} with $\tilde f^{abc}$ color factors
at each diagram vertex.  The planar integrands are given in \eqn{eq:PlanarIntegrand}
while the nonplanar integrands are given in
\eqns{eq:symIntegrand1}{eq:symIntegrand1}.  The $A^\tree$'s are
color-ordered tree amplitudes of $\mathcal N = 4$ super-Yang--Mills
theory, with the indicated ordering of legs.  The overall sum is over all 24 permutations of the
external legs; the permutations act on the external color,
polarization and momentum labels.  The form in
Eq.\ \eqref{eq:TwoLoopAmplitude} differs from the one
originally given in Ref.~\cite{BDDPR} by terms that vanish via the
color Jacobi identity.  In the original form, the individual nonplanar
integrals do not reflect the analytic properties of the final
amplitude, such as having only logarithmic singularities and no poles
at infinity.

In order to understand the transformation properties we divide 
the amplitude into sectors determined by the tree amplitude prefactors.
These tree amplitudes have differing overall weights under the  
transformations, which are easy to determine using the identities~\cite{BCJ},
\begin{equation}
s t A^\tree (1,2,3,4) = s u A^\tree(1,2,4,3) = tu A^\tree(1,3,2,4)\,.
\end{equation}
From here we can see that the tree amplitudes transform with different 
overall weights under \eqn{eq:NonplanarDoubleBox}, 
\begin{equation}
\Delta \biggl(\frac{A^\tree(1,2,3,4)}{A^\tree(1,2,4,3)} \biggr)
 = \Delta \Bigl(\frac{t}{u}\Bigr) = - \frac{st}{2u} \,,
\end{equation}
where we used \eqns{eq:npdbox-DeltaST}{eq:npdbox-DeltaU}.

In \eqn{eq:TwoLoopAmplitude}, the coefficient of each tree amplitude
factor is invariant under the four-dimensional symmetry.  In
Refs.~\cite{NonplanarAmplituhedron}, the ordering of the tree factors
were chosen to adjust the factors of $s$, $t$, and $u$ so that the
remaining integrals have unit leading singularities. Not surprisingly,
these factors are exactly what is needed to make the coefficient of
each tree invariant under the four-dimensional symmetry.

As a side note, we can adjust the transformations in
each sector so that a uniform transformation is applied to the
external momenta in all sectors of the amplitude.  In doing so, the
transformations on internal momenta necessarily differ in the various
sectors, as expected from the fact that there is no uniform sets of
momenta or dual variables in the nonplanar sector. This 
may be accomplished by adjusting Lorentz and scaling 
transformations. However, since the different sectors transform
with a different weight there is no need to do this.

The transformations described in the  previous
section can be taken as a direct analog of dual conformal symmetry of planar
$\mathcal N=4$ super-Yang--Mills theory, but applicable to the
nonplanar sector as well. Like dual conformal symmetry in the planar case, the infinitesimal generators of the new symmetries can be identified as polynomial tangent vectors of unitarity cut surfaces.

This opens the possibility of finding numerators of higher-loop
integrals with desired properties of having simple analytic
properties and associated DEs, not by detailed studies of
the singularity structure of the integrands~\cite{HennDifferentialEqs,
  TwoLoopN4Nonplanar, NonplanarAmplituhedron}, but by demanding
that given integrands be invariant under symmetries analogous to dual
conformal symmetry.  For nonplanar diagrams that can be obtained from
a planar one by a single replacement of a box subdiagram by a crossed
box, as in \fig{fig:CrossedMultiloop}, the obvious candidate
transformations follow those described in the previous section.  It
would be very interesting to systematically study these cases, as well
as ones with multiple twists.  We expect such integrals to be direct
building blocks for nonplanar $\mathcal N = 4$ super-Yang--Mills
amplitudes. More generally, it seems likely that a symmetry along the
lines described here is responsible for the simple analytic
properties~\cite{TwoLoopN4Nonplanar,NonplanarAmplituhedron} of general
nonplanar amplitudes at any loop order.


\section{IBP-generating vectors and Landau equations}
\label{sec:Landau}

In previous sections we obtained IBP-generating vectors from the point
of view of conformal symmetries. In the following we will see that the
symmetry generators occur with vanishing or degenerate mass
configurations. We will show that the defining
relations of the vectors connect directly to Landau equations, which
determine the presence/absence of the exceptional vectors.

For simplicity we consider $N$-point one-loop integrals in the
following.  When IBP-generating vectors are expressed in dual
coordinates, they fulfill a number of consistency
requirements~\cite{CaronHuotEmbedding}:
(i) They are tangent to a quadric $(YY)=0$, which defines the integration
contour, 
(ii) being vector fields in projective space, they are homogeneous in the
variable $Y$, 
(iii) and vectors are defined modulo vectors proportional to $Y^a$ which induce
a $GL(1)$ scaling and give trivial identities.  

Maintaining propagator powers of the integrals, yields the further conditions,
\begin{eqnarray}
    V^a\frac{\partial}{\partial Y^a} \, (X_i Y) = {\cal W}_i (X_i Y) \,, \quad \quad 1 \leq i \leq N\,,
\end{eqnarray}
and in total we have the linear relations,
\begin{eqnarray} \label{ibp-system}
        (X_i V)  - (X_iY) {\cal W}_i =0\,, \quad \quad
         2 (YV)  = 0\,,  
\end{eqnarray}
for the data of the IBP-generating vector $V^a$ and the weight functions ${\cal W}_k$.
The above relations hold up to terms proportional to $(YY)$. 

It is convenient to write Eqs.~(\ref{ibp-system}) in matrix form in terms of
the vectors in the problem. To this end we expand the IBP-generating vector in
terms of the infinity point, $X_j$ and transverse points $N_j$, the number of
which varies in dimensional regularization, but do not include $Y$, i.e. 
\begin{equation}
V=\biggl(\sum_{j=1}^N X_j v^j \biggr) +  I v^{N+1} + \biggl(\sum_{k=N+2}^{d+2} N_k v^k \biggr)\,.
\end{equation}
We obtain the matrix,
\begin{eqnarray}
\hskip -.3 cm 
    C={ \left(\begin{array}{cccccccccc}  
        (X_1X_1) &  (X_1X_2) & \cdots &  1  & 0          &  0           & \cdots & - (X_1Y)  &       0   &  0 \\
        (X_2X_1) &  (X_2X_2) & \cdots &  1  & 0          &  0           & \cdots & 0         & - (X_2Y)  &  0  \\
        \cdots    &            &        &     &          &              & &      &           &     \\
        (X_NX_1) &  (X_NX_2) & \cdots &  1  & 0          &  0           & \cdots & 0         &    0      &  - (X_N Y)   \\
        2(X_1 Y)  &   2(X_2 Y) & \cdots &  2 (YI)  & 2(YN_1)   &  2(YN_2) & \cdots & 0         &  \cdots   &  0    
    \end{array}\right) }\,. \nn \\
\label{CMatrix}
\end{eqnarray}
which has been simplified using our gauge choice, $(X_iI) = 1$ of the
dual conformal embedding, and using the transversality
$(N_iX_j)=(N_iI)=0$.  This matrix acts on the column vector $(v^i, {\cal
  W}_k)$.  The entries of the matrix $C$ are monomials of either degree 0 or degree 1 in $Y$, which leads to, in an approach using computational algebraic geometry (not covered here), simple syzygy equations that could be solved efficiently.
The top-left $(N+1)\times (N+1)$ sub-block of the above matrix is
closely related to the ``embedding space Gram matrix'' used in Section
\ref{sec:triangle}. The top-left $N\times N$ sub-block defines the
Cayley determinant
$\Delta_C=\det(\{(X_iX_j)\}_{i,j=1,N})$~\cite{AbreuBrittoEtAl}.

From \eqn{ibp-system} we see that the IBP-generating vectors $(v^i, {\cal W}_k )$ correspond
to the kernel of the matrix $C$.  The generic vectors in the kernel of the
matrix $C$ are obtained by applying Cramer's rule, with vector components being
appropriate minors of the linear system. Cramer's rule gives,
\begin{eqnarray} 
    c_{a_1} ( a_2 , a_3 , \cdots , a_{N+2} ) -  
    c_{a_2} ( a_1 , a_3,       \cdots , a_{N+2} ) + \cdots + 
    (-1)^{N+1} c_{a_{N+2}} ( a_1 , a_2 , \cdots , a_{N+1} ) =0 \,. \nn \\
\label{Cramer}
\end{eqnarray}
The expressions $c_{a_i}$ stand for columns of the original matrix
$C$ and $(a_1,a_2, \cdots )$ stand for the leading minors of the
matrix $C$ given by the determinants of the matrices composed of the
columns $ (a_1,a_2, \cdots ) \equiv \det (c_{a_1}, c_{a_2}, \cdots)$.\footnote{This notation has been used in e.g.\ Ref.\cite{DLogForms} in the literature.}
\Eqn{Cramer} is a statement that $N+2$ vectors in $N+1$ dimensions are
necessarily linearly dependent.

Interesting solutions are the ones which do not vanish on the cuts,
i.e. on the surfaces defined by $(X_iY)=0$.
The respective IBP-generating vectors are then obtained by choosing $(N+1)$
columns, where we distinguish the following subclasses using the entries of the last row:
\begin{enumerate}
\item Two $(YN_i)$ factors and no $(YI)$ factor,
\begin{eqnarray}
    V=\Delta_C\, (YN_{[i|})N_{|j]}\,.
\end{eqnarray}
We can factor out $\Delta_C$ from the above expression, so we still use the transverse-space rotation vector that is left, when $\Delta_C=0$.
\item  One $(YN_i)$ factor and one $(YI)$ factor,
\begin{eqnarray}
    V&=&\Delta_C\,(YN_{k})I  
        -(1,2,\cdots , N+1) N_k\nn \\
        && + \sum_{j=1}^N (-1)^{N+j+1} X_j (1,2,\cdots, \widehat{j},\cdots , N+1, N+1+k)\,,
\end{eqnarray}
which matches the vector (\ref{eq:UtildeINk}) up to normalization by
$\Delta_C$. The hat-symbol, $\widehat{j}$ indicates that the $j^{th}$ column is omitted.
\end{enumerate}
The above are the generic IBP-generating vectors. We have omitted the vectors that vanish on the maximal cut.

In addition, exceptional vectors appear for particular kinematic
configurations. For one-loop integrals we consider the case of a vanishing
Cayley determinant, $\Delta_C=0$. A maximal-cut analysis predicts a scaling
vector in addition to rotation vectors~\cite{ItaProceeding, ItaIBP}.  The
following construction shows how the vector arises from Lie-algebra
generators acting on the Gram-matrix $\{(X_iX_j)\}_{i,j=1,N}$. We consider a
generator with the first $(N+1)$ components non-vanishing and allow for
non-vanishing ${\cal W}$'s. The generator has to give zero when contracted with
either of the rows. Starting with the last row we find the general form,
\begin{eqnarray} v^i=\sum_{b=1}^{N+1}\omega^{ij} C_{N+1,j}\,, \end{eqnarray}
and an antisymmetric matrix $\omega^{ij}$. We attempt to find a solution of minimal
degree, thus we assume $\omega^{ij}$ to be constant. In the language of
computational algebraic
geometry, this is the general form to parametrize the principal syzygies of the
first $(N+1)$ elements of the last row. 

An interesting consistency condition appears when considering the
$(YI)$ terms.  The argument uses that $(YI)$ enters through the last
column of $\omega^{ij}$, with $\omega^{N+1,N+1}=0$ from
antisymmetry. Thus, in order for the $(YI)$ to cancel for each
contraction, we require that the column vector $\omega^{N+1,j}$ is in
the kernel of the $\{(X_iX_j)\}_{i,j=1,N}$ block of the linear system,
confirming that the condition $\Delta_C=0$ is necessary for these
types of vectors.

This brings us to a close connection between the $C$ matrix in
\eqn{CMatrix} and the Landau
equations~\cite{Landau,SMaxrixBooks,AbreuBrittoEtAl}, whose
embedding-space version is,
\begin{eqnarray}
  \sum_{i=1}^N \alpha_i X_i + Y =0 \,, \hskip 1 cm \alpha_j (X_j Y) =0\,.
\label{landau}
\end{eqnarray}
(For simplicity we omitted the extension of the Landau equations which
in addition capture the singularities at infinity.) The leading Landau
singularity appears for all propagators vanishing, $(X_j Y)=0$, and
requiring that the points $X_i$ and $Y$ must be linearly
dependent. The equations are easiest analyzed when contracted with an
independent set of points $\{Y,X_j,N_k,I\}$. The non-trivial
conditions then are, setting $(X_j Y)=0$,
\begin{eqnarray} \label{eq:landau-eqns} (N_iY)=0\,, \hskip 0.8 cm \sum_{j=1}^N \alpha_j
    (X_jX_k)=0 \,, \hskip 0.8 cm \left( \sum_{i=1}^N \alpha_i (X_iI) \right)+(YI)=0\,.
\end{eqnarray}
Remarkably, this can be re-written,  using the transpose of the $C$ matrix, as
\begin{equation}
C^T \begin{pmatrix} \alpha_1 \\ \vdots \\ \alpha_N \\ 1 \end{pmatrix} = 0 \, .
\end{equation}
The conditions \eqn{eq:landau-eqns} in turn imply $\Delta_C=0$ while the overall normalization of
the $\alpha$'s is fixed by the last equations in (\ref{eq:landau-eqns}) using the
gauge conditions $(X_iI)=(YI)=1$. In this way the Landau conditions
(\ref{landau}) determining a singular point of the on-shell surface relate to
the solutions of the IBP-generator equations (\ref{ibp-system}).

Thus, we have given the form of the generic IBP-generating vectors in dual coordinates. In addition, we showed that exceptional vectors appear
for degenerate kinematic configurations.  We pointed out the close relation between
the defining equations for unitarity-compatible vectors fields and the Landau
equations. In this way, Landau equations 
signal the appearance of exceptional IBP-generating vectors.
This connection between Landau equations and IBP relations might be expected, given
that Landau equations signal a singular kinematic configuration in which the set of
master integrals is reduced and additional integral relations appear.

The minor construction generalizes naturally to higher-loop computations, so
that minors of the linear system of the IBP-generator equations play a central role
for constructing IBP-vectors~\cite{ItaIBP}. Additional investigations are needed to systematize the description of exceptional vectors at higher loops.  Here the on-shell surfaces have a
more complicated topology, with the characteristic sizes determined by
combinations of the kinematic invariants (like the Cayley determinant $\Delta_C$).

\section{Conclusion}
\label{sec:Conclusion}

In this paper we studied hidden symmetries of $\mathcal N = 4$
super-Yang--Mills theory as a means for generating compact integration-by-parts (IBP)
relations~\cite{IBP} and differential equations (DEs)~\cite{DEs} for loop
integrals encountered in generic theories.  For the planar case, the
hidden symmetry is the well-studied dual conformal
symmetry~\cite{DCI}.  By exploiting the connection between dual
conformal symmetry and polynomial tangent vectors of unitarity cut
surfaces, we were able to find an analogous symmetry for the nonplanar
sector of the two-loop four-point amplitude as well.  Besides being
useful for generating IBP relations and DEs, this
points to the exciting possibility that dual conformal symmetry can
be generalized to the nonplanar sector of $\mathcal N = 4$
super-Yang--Mills theory.

Dual conformal transformations and their nonplanar analogs have the
important property that they do not increase propagator powers,
resulting in IBP relations and DEs that are
naturally compatible with unitarity~\cite{KosowerNoDouble}.  Such IBP
relations had been previously described using computational algebraic
geometry~\cite{KosowerNoDouble, Schabinger, LarsenZhang}.  Our
approach, based on exploiting hidden symmetries, provides new analytic
insights and on the practical side gives compact expressions for the
IBP-generating vectors and DEs.

In describing the symmetries we found it useful to work with both
``direct'' dual conformal transformations in $d$ dimensions and the
embedding formalism~\cite{SimmonsDuffin}, which linearizes the
transformations by going to $(d+2)$ dimensions.  The Gram matrix
defined in the embedding formalism also clarifies the connection to
Landau equations.

To illustrate these ideas, we presented a variety of examples at one
and two loops. With up to four massless legs and a small number of
mass parameters, it is straightforward to find several dual conformal
transformations which leave the external momenta invariant, and lead
to a sufficient number of IBP relations to solve generic cases. For
example, the dual conformal transformations generate a complete set of
IBP relations for the planar two-loop double box integral. We
also studied a five-point example, namely the planar penta-box
integral. In this case, we need additional IBP-generating vectors from
combining separate conformal transformations for the left loop and
right loop, generalizing the strategy of Ref.~\cite{ItaIBP}. These
additional vectors still have a simple analytic form. For
illustration, we also looked at a simpler three-point nonplanar
integral, and  obtained IBP relations that reduce all
integrals to top-level master integral and daughter integrals.

We also described DEs, where the integrals do not
have raised propagator powers, for both planar and nonplanar cases
that arise when external momenta are allowed to change under the
transformation.  For one- and two-loop integrals with appropriately
chosen numerators that make the transformation weights cancel in four
dimensions, the method directly gives a DEs where
the right hand side is proportional to the dimensional regularization
parameter $\epsilon$~\cite{HennDifferentialEqs}.  This holds 
before IBP reduction to a basis of master integrals, because the
equations follow from a symmetry that is exact in four dimensions.
For massless kinematics, the method is applicable with up to five
external legs. At higher points, when nontrivial conformal cross
ratios are present, the method generates a subset of the DEs. 

Our results point to promising directions for future studies.  In
various one- and two-loop examples we showed the utility of dual
conformal invariance for generating both IBP relations and
DEs, as well as presented a nonplanar symmetry
analogous to dual conformal symmetry.  An obvious direction for future
studies is to try to generalize this to arbitrary loop orders and for
any number of external legs.  The unitarity-compatible IBP-generating
vectors and DEs constructed via dual conformal
symmetry and its generalizations are particularly simple, making it
desirable to extend these ideas as widely as possible.  The ability to
generate relatively simple DEs becomes especially
attractive when existing methods suffer from computational bottlenecks
that occur in more complicated cases.  It is also worth studying
whether the compact expressions generated from our symmetry
considerations can improve computational efficiency in numerical
unitarity approaches at two loops and beyond~\cite{BH4DGluon}.  We
also noted an important connection between IBP-generating vectors
and Landau equations, which would be interesting to pursue.

On the more formal side, we know that dual conformal
symmetry~\cite{DCI} strongly restricts the analytic properties of the
planar sector of $\NeqFour$ super-Yang--Mills theory. In particular,
the integrands have no double poles or poles at
infinity~\cite{DLogForms}. These analytic properties also appear to
carry over to the nonplanar
sector~\cite{TwoLoopN4Nonplanar,NonplanarAmplituhedron}.  Here we took
initial steps to identify a symmetry that can explain this.  We
explicitly constructed a symmetry of the nonplanar two-loop four-point
$\NeqFour$ amplitude, and used it to construct a differential equation
for determining its value. As in the planar case, the symmetry is
intimately connected to polynomial tangent vectors of unitarity cut
surfaces. As for dual conformal invariance the symmetry is anomalous
due to infrared singularities.  We noted that for the class of
integrals with a single crossed box and the remaining part planar, the
symmetry extends straightforwardly to all loop orders with an
arbitrary number of external legs.  An important next step would be to
extend this to more general nonplanar cases. 

We look forward to exploring these ideas for simplifying computations
of multi-loop integrals needed for scattering cross sections at
particle colliders, as well as for understanding hidden symmetries of
the nonplanar sector of $\mathcal N = 4$ super-Yang--Mills theory. 
These two issues are intertwined, as we found here.

\section*{Acknowledgments}
We thank Samuel Abreu, Julio Parra-Martinez, Ben Page, Chia-Hsien Shen, Jaroslav Trnka
and Yang Zhang for many useful and interesting discussions.  This research 
is supported by the Department of Energy under Award Numbers
DE-SC0009937.  H.I.'s work is supported by a Marie Sk{\l}odowska-Curie
Action Career-Integration Grant PCIG12-GA-2012-334228 of the European
Union. This research is supported by the Munich Institute for Astro-
and Particle Physics (MIAPP) of the DFG cluster of excellence `Origin
and Structure of the Universe'.

\appendix

\section{Sub-loop IBP-generating vectors for the penta-box}
\label{sec:pentaboxDetails}
In this appendix, we tabulate the antisymmetric matrices in Eq.\ \eqref{eq:pentaBoxSubLoopVecs} of Subsection \ref{subsec:pentabox}, which parametrize conformal transformations that leave a subset of external momenta invariant. The matrices are,
\begin{align}
\omega_{(1)} &= \begin{pmatrix}
  0 & s_{34} & s_{51} & s_{34}-s_{51} & s_{23} & s_{23} s_{34} \\
 -s_{34} & 0 & 0 & -s_{34} & s_{12}-s_{45} & -s_{34} s_{45} \\
 -s_{51} & 0 & 0 & -s_{51} & s_{23}-s_{45} & -s_{45} s_{51} \\
 -s_{34}+s_{51} & s_{34} & s_{51} & 0 & s_{12} & s_{12} s_{51} \\
 -s_{23} & -s_{12}+s_{45} & -s_{23}+s_{45} & -s_{12} & 0 & -s_{12} s_{23} \\
 -s_{23} s_{34} & s_{34} s_{45} & s_{45} s_{51} & -s_{12} s_{51} & s_{12} s_{23} & 0
 \end{pmatrix}, \nn \\
\omega_{(2a)} &= \begin{pmatrix}
 0 & 0 & 0 & 0 & 0 & 0 \\
 0 & 0 & 0 & 0 & -s_{45} & 0 \\
 0 & 0 & 0 & 0 & 0 & 0 \\
 0 & 0 & 0 & 0 & -s_{23} & 0 \\
 0 & s_{45} & 0 & s_{23} & 0 & s_{23} s_{45} \\
 0 & 0 & 0 & 0 & -s_{23} s_{45} & 0 \\
 \end{pmatrix}, \nn \\
\omega_{(2b)} &= \begin{pmatrix}
 0 & -s_{34} & -s_{23}+s_{51} & 0 & 0 & -s_{23} s_{34} \\
 s_{34} & 0 & s_{45} & -s_{12}+s_{34} & 0 & s_{34} s_{45} \\
 s_{23}-s_{51} & -s_{45} & 0 & -s_{51} & 0 & -s_{45} s_{51} \\
 0 & s_{12}-s_{34} & s_{51} & 0 & 0 & s_{12} s_{51} \\
 0 & 0 & 0 & 0 & 0 & 0 \\
 s_{23} s_{34} & -s_{34} s_{45} & s_{45} s_{51} & -s_{12} s_{51} & 0 & 0
 \end{pmatrix}, \nn \\
\omega_{(2c)} &= \begin{pmatrix}
 0 & 0 & 0 & 0 & s_{23} & 0 \\
 0 & 0 & 0 & 0 & -s_{45} & 0 \\
 0 & 0 & 0 & 0 & 0 & 0 \\
 0 & 0 & 0 & 0 & 0 & 0 \\
 -s_{23} & s_{45} & 0 & 0 & 0 & 0 \\
 0 & 0 & 0 & 0 & 0 & 0
 \end{pmatrix} .
\end{align}


\end{document}